\documentclass{tlp}
\usepackage{times}
\usepackage{amssymb}
\usepackage{amsmath}
\usepackage{stmaryrd}
\usepackage{cdsty}
\usepackage{ifthen}
\usepackage{hyperref}
\usepackage{color}
\usepackage[all]{xy}

\newcommand{\sTrue}{\top}              
\newcommand{\True}{\sTrue}

\newcommand{\sAnd}{\land}             
\renewcommand{\And}{\, \sAnd \,}

\newcommand{\sImp}{\supset}            
\newcommand{\Imp}{\sImp}
\newcommand{\sPim}{\subset}            
\newcommand{\Pim}{\sPim}
\newcommand{\sE}{\exists}              
\newcommand{\E}[2]{\sE #1.\, #2}
\newcommand{\sA}{\forall}              
\newcommand{\A}[2]{\sA #1.\, #2}

\newcommand{\cRule}[3]
 {\ianc{#1}{#2}{\mbox{$\bf \scriptstyle #3$}}}

\newcommand{\qq}
 {\quad}

\newcommand{\rname}[1]{\ensuremath{\mathbf{#1}}}

\newcommand{\G}{\Gamma}

\renewcommand{\S}{\Sigma}

\newcommand{\unid}[2]                             
 {#1 \stackrel{u}{\longrightarrow} \, #2}
\newcommand{\immd}[3][a]                          
 {#2 \stackrel{u}{\longrightarrow} \, #3 \, \gg \, #1}

\newcommand{\bigLam}[2][\alpha]
 {\Lambda #1.\, #2}

\newcommand{\cmpProg}[2]                          
 {#1 \, \gg \, #2}
\newcommand{\cmpClause}[3][\alpha]                
 {#2 \, \gg \, #1 \, \backslash \, #3}
\newcommand{\cmpGoal}[2]                          
 {#1 \, \gg \, #2}

\newcommand{\gResA}[3][c_0]                        
 {#2 \stackrel{#1}{\longrightarrow} \, #3}
\newcommand{\rResA}[3][c_0]                        
 {#2 \stackrel{#1}{\longrightarrow} \, #3}

\newcommand{\gResB}[2]{\gResA[c_1]{#1}{#2}}          
\newcommand{\rResB}[2]{\rResA[c_1]{#1}{#2}}          
\newcommand{\cResB}[3][a]
 {#2 \stackrel{c_1}{\longrightarrow} \, #3 \, \gg \, #1}

\newcommand{\cmpProgB}[2]
 {#1 \, \gg \, #2}
\newcommand{\cmpClauseB}[3][\CC]
 {#2 \, \gg \, #1 \, \backslash \, #3}
\newcommand{\cmpGoalB}[2]
 {#1 \, \gg \, #2}
\newcommand{\cmpHeadB}[4]
 {#1 \vdash \cmpClauseB[#3]{#2}{#4}}

\newcommand{\mResC}[2]{\gResA[c_2]{#1}{#2}}          
\newcommand{\fResC}[2]{\gResA[c_{2f}]{#1}{#2}}          
\newcommand{\gResC}[2]{\gResA[c_2]{#1}{#2}}          
\newcommand{\rResC}[2]{\rResA[c_2]{#1}{#2}}          
\newcommand{\cResC}[3][a]
 {#2 \stackrel{c_2}{\longrightarrow} \, #3 \, \gg \, #1}

\newcommand{\cmpProgC}[2]
 {#1 \, \gg \, #2}
\newcommand{\cmpClauseC}[4][\CC]
 {#2 \, \gg \, #1 \, \backslash \, #3 \nnwarrow #4}
\newcommand{\cmpGoalC}[2]
 {#1 \, \gg \, #2}
\newcommand{\cmpHeadC}[5]
 {#1 \vdash #2 \,\gg\, #3 \ssearrow #4 \nnwarrow #5}
\newcommand{\cmpAtmC}[4]
 {#1 \vdash #2 \,\gg\, #3 \,\backslash\, #4}

\newcommand{\mynote}[2][5cm]
 {\marginpar{ \parbox{#1}{\small #2}}}
\renewcommand{\mynote}[1]{}

\newcommand{\uObj}[4][\S]                         
 {#2 \stackrel{u}{\longrightarrow}_{#1}
     \, #3 : #4}
\newcommand{\iObj}[6][\S]                         
 {#2 \stackrel{u}{\longrightarrow}_{#1}
     \, #3 : #4 \, \gg \, #5 : #6}

\newcommand{\sEq}{\stackrel{.}{=}}
\newcommand{\Eq}[2]
 {#1 \sEq #2}

\newcommand{\rObj}[4][\S]                         
 {#2 \stackrel{r}{\longrightarrow}_{#1}
     \, #3 : #4}
\newcommand{\Dec}[6]                              
 {#1 : #2 \, \gg \, #3 : #4
          \, \backslash \, #5 : #6}

\newcommand{\ms}[1]{\mathsf{#1}}
\renewcommand{\vec}[1]{\underline{#1}}

\newcommand{\IN}[1]{\check{#1}}
\newcommand{\OT}[1]{\hat{#1}}
\newcommand{\Ivec}[1]{\vec{\IN{#1}}}
\newcommand{\Ovec}[1]{\vec{\OT{#1}}}

\newcommand{\Match}[2]{#1 =: #2}
\newcommand{\Assg}[2]{#1 := #2}

\newcommand{\C}{\tilde{C}}
\newcommand{\R}{\tilde{R}}
\newcommand{\CC}{\mathcal{C}}
\newcommand{\GG}{\mathcal{F}}

\renewcommand{\L}{\mathcal{L}}
\newcommand{\Ls}{\L^s}
\newcommand{\Lc}{\L^c_0}
\newcommand{\LcB}{\L^c_1}
\newcommand{\LcC}{\L^c_2}

\begin{document}
\bibliographystyle{acmtrans}

\title{An Improved Proof-Theoretic Compilation of Logic Programs}

\author[I. Cervesato]
{Iliano Cervesato \\
Department of Computer Science\\
Carnegie Mellon University\\
E-mail: iliano@cmu.edu
}

\pagerange{\pageref{firstpage}--\pageref{lastpage}}
\volume{\textbf{10} (3):}
\jdate{TBA}
\setcounter{page}{1}
\pubyear{TBA}

\maketitle

\label{firstpage}

\begin{abstract}
  In prior work, we showed that logic programming compilation can be given a
  proof-theoretic justification for generic abstract logic programming
  languages, and demonstrated this technique in the case of hereditary Harrop
  formulas and their linear variant.  Compiled clauses were themselves logic
  formulas except for the presence of a second-order abstraction over the
  atomic goals matching their head.  In this paper, we revisit our previous
  results into a more detailed and fully logical justification that does away
  with this spurious abstraction.  We then refine the resulting technique to
  support well-moded programs efficiently.

  \emph{To appear in Theory and Practice of Logic Programming.}
\end{abstract}

\section{Introduction}
\label{sec:intro}

In~\cite{Cervesato98jicslp}, we presented a general methodology for developing
a compiler and associated intermediate language for any abstract logic
programming language (ALPL) \cite{Miller91apal} that satisfies some basic
proof-theoretic properties.  We applied it abstractly to the language of
hereditary Harrop formulas and its linear variant, and also based the concrete
implementations of the Twelf~\cite{Pfenning99cade} and LLF~\cite{ic02} systems
directly on it.  This methodology identified right sequent rules that behave
like the left rules that can appear in a uniform proof and used the
corresponding connectives as the compilation targets of the constructs in
program clauses.  The intermediate language was therefore just another ALPL
and its abstract machine relied on proof-search, like the source ALPL\@.
Because the transformation was based on the proof-theoretic duality between
left and right rules, proving the correctness of the compilation process
amounted to a simple induction.  Finally, for Horn clauses the connectives in
the target ALPL corresponded to key instructions in the Warren Abstract
Machine (WAM)~\cite{Warren83tr}.  The WAM is an essential component of
commercial Prolog systems since many compiled programs run over an order of
magnitude faster than when interpreted.

Up to then, the notoriously procedural instruction set of the WAM was regarded
as a wondrous piece of engineering without any logical status, in sharp
contrast with the deep logical roots of Prolog.  In the words
of~\cite{Boerger95fmpa} ``[the WAM] resembles an intricate puzzle, whose many
pieces fit tightly together in a miraculous way''.  As a result, understanding
it was complex in spite of the availability of excellent
tutorials~\cite{Aitkaci91book}, proving its correctness was a formidable
task~\cite{Boerger95fmpa,Russinoff92jlp}, and adapting it to other logic
programming languages a major endeavor --- it was done for \emph{CLP$({\cal
    R})$}~\cite{Jaffar92pldi} and
\emph{$\lambda$Prolog}~\cite{nadathur99cade}.  By contrast, the methodology
in~\cite{Cervesato98jicslp} is simple, (mostly) logic-based, easily
verifiable, and of general applicability.

The technique in~\cite{Cervesato98jicslp} had however one blemish: it made use
of equality over atomic formulas together with a second-order binder over
atomic goals, which lacked logical status.  In this paper, we
remedy this drawback by carefully massaging the head of clauses.  This allows
us to replace those constructs with term-level equality and regular universal
quantifications over the arguments of a clause head.  The result is an
improved proof-theoretic account of compilation for logic programs that sits
squarely within logic.  It also opens the doors to specializing the
compilation process to well-moded programs, which brings out the potential of
doing away with unification in favor of matching, a more efficient operation
in many languages.  We present these results for the language of hereditary
Harrop formulas and only at the highest level of abstraction.  Just
like~\cite{Cervesato98jicslp}, they are however general, both in terms of the
source ALPL and of the level of the abstraction considered.  We are indeed in
the process of using them to implement a compiler for
CLF~\cite{cmu-cs-02-101,cmu-cs-02-102}, a higher-order concurrent linear logic
programming language that combines backward and forward chaining.

The paper is organized as follows: Section~\ref{sec:pil0} recalls the
compilation process of~\cite{Cervesato98jicslp}.  In Section~\ref{sec:pil1},
we present our improved compilation process.  In Section~\ref{sec:pil2}, we
refine it to support moded programs.  We lay out future developments in
Sections~\ref{sec:larger-languages} and~\ref{sec:future}.

\section{Background and Recap}
\label{sec:pil0}

In this section, we recall the compilation process presented
in~\cite{Cervesato98jicslp}.  For succinctness, we focus on a smaller source
language --- it corresponds to the language underlying the Twelf
system~\cite{Pfenning99cade}, on which this technique was first used.  We will
comment on larger languages, including those examined
in~\cite{Cervesato98jicslp}, in Section~\ref{sec:larger-languages}.

\subsection{Source Language}
\label{sec:pil0-source}

We take the language freely generated from atomic propositions ($a$),
intuitionistic implication ($\Imp$) and universal quantification ($\forall$)
as our source language.  We expand the open-ended atomic propositions
of~\cite{Cervesato98jicslp}, into a \emph{predicate symbol} $p$ followed by
zero or more terms $t$.  A program is a sequence of closed formulas.  This
language, which we call $\Ls$, is given by the following grammar:

\medskip
\noindent
{\arraycolsep=0.25em
\newcommand{\sep}{\hspace{0.5em}|\hspace{0.5em}}
$\begin{array}[t]{@{}r@{\hspace{0.8em}} cr l@{}}
  \mbox{\emph{Formulas:}}
& A & ::= & a
      \sep  A_1 \Imp A_2
      \sep  \A{x}{A}
\\ \mbox{\emph{Atoms:}}
& a & ::= & p
      \sep a\:t
\end{array}$
\hfill
$\begin{array}[t]{@{}r@{\hspace{0.8em}} cr l@{}}
 \mbox{\emph{Programs:}}
& \G & ::= & \cdot
       \sep  \G, A
\end{array}$
\hspace*{1em}}

\medskip

\noindent
As in~\cite{Cervesato98jicslp}, we leave the language of terms open, but
require that it be predicative (substituting a term for a variable cannot
alter the outer structure of a formula).  We will often write an atom $a$ as
$p\;\vec{t}$, where $p$ is its predicate symbol and $\vec{t}$ is the sequence
of terms it is applied to.
We implicitly assume that a predicate symbol is consistently applied to the
same number of terms throughout a program --- its arity.
We write $[t'/x]t$ (resp.\ $[t'/x]A$) for the capture-avoiding substitution of
term $t'$ for all free occurrences of variable $x$ in term $t$ (resp.\ in
formula $A$).  Simultaneous substitution is denoted $[\vec{t'}/\vec{x}]t$ and
$[\vec{t'}/\vec{x}]A$.

$\Ls$ is an abstract logic programming language~\cite{Miller91apal} and, for
appropriate choices of the term language, has indeed the same expressive power
as $\lambda$Prolog~\cite{Miller86iclp} or Twelf~\cite{Pfenning99cade}.  It
differs from the first language discussed in~\cite{Cervesato98jicslp} for the
omission of conjunction and truth (see Section~\ref{sec:larger-languages}).

The operational semantics of $\Ls$ is given by the two judgments
$$
\begin{array}{l@{\hspace{1.5em}}p{16em}}
   \unid{\G}{A}
 & \emph{$A$ is uniformly provable from $\G$}
\\ \immd{\G}{A}
 & \emph{$a$ is immediately entailed by $A$ in $\G$}
\end{array}
$$
Their defining rules, given in Figure~\ref{fig:pil0-uniform}, produce uniform
proofs~\cite{Miller91apal}: the uniform provability judgment includes the
right sequent rules for $\Ls$ and, once the goal is atomic, rule
\rname{u\_atm} calls the immediate entailment judgment, which focuses on a
program formula $A$ and decomposes it as prescribed by the left sequent rules.
This strategy is complete with respect to the traditional sequent rules of
this logic~\cite{Miller91apal}.  From a logic programming perspective, the
connectives appearing in the goal --- handled by right rules --- are search
directives, while the left rules carry out a run-time preparatory phase.

\begin{figure}[t]
\begin{center}
\leavevmode
\newcommand{\rsp}{1.8ex}
\newcommand{\rspl}{-.5ex}
\newcommand{\trule}[1][-1.5ex]{\rule[#1]{0em}{0ex}}
\newcommand{\hhfill}{\hfill}
\fboxsep=0pt

\noindent
\fbox{\scriptsize
$\begin{array}{@{\hfill}c@{\hfill}}
\makebox[\textwidth]{}\\[-2.5ex]
\multicolumn{1}{l}{\bf \scriptstyle Uniform \; provability\trule}
\\
  \cRule
   {\immd{\G, A, \G'}{A}}
   {\unid{\G, A, \G'}{a}}
   {u\_atm}
\hhfill
  \cRule
   {\unid{\G, A_1}{A_2}}
   {\unid{\G}{A_1 \Imp A_2}}
   {u\_imp}
\hhfill
  \cRule
   {c \; \mbox{\em ``new''}
    \qq
    \unid{\G}{[c/x]A}}
   {\unid{\G}{\A{x}{A}}}
   {u\_all}
\\[\rspl]
\hline\\[-4.5ex]
\multicolumn{1}{l}{\bf \scriptstyle Immediate \; entailment\trule}
\\[-1.0ex]
  \cRule
   {}
   {\immd{\G}{a}}
   {i\_atm}
\hhfill
  \cRule
   {\immd{\G}{A_1}
    \qq
    \unid{\G}{A_2}}
   {\immd{\G}{A_2 \Imp A_1}}
   {i\_imp}
\hhfill
  \cRule
   {\immd{\G}{[t/x]A}}
   {\immd{\G}{\A{x}{A}}}
   {i\_all}
\\[-1ex]\relax
\end{array}$}

\caption{Uniform Deduction System for $\Ls$.}
\label{fig:pil0-uniform}
\end{center}
\end{figure}

\subsection{Target Language}
\label{sec:pil0-target}

In~\cite{Cervesato98jicslp}, the target language of the compilation process
distinguished compiled goals ($G$) from compiled clauses ($C$).  A compiled
goal was either an atomic proposition, or a hypothetical goal (a goal to be
solved in the presence of an additional clause) or a universal goal (a goal to
be solved in the presence of a new constant).  A compiled clause had the form
$\bigLam{C}$, where the second-order variable $\alpha$ stood for the atomic
goal to be resolved against the present clause, while $C$ could either match
$\alpha$ with the head $a$ of this clause ($\Eq{a}{\alpha}$), invoke a goal
($C \And G$), or request that a variable $x$ be instantiated with a term
($\E{x}{C}$).  A compiled program $\Psi$ was then a sequence of compiled
clauses.  The grammar for the resulting language, which we call $\Lc$, is as
follows:

\medskip
\noindent
{\arraycolsep=0.25em
\newcommand{\sep}{\hspace{0.5em}|\hspace{0.5em}}
$\begin{array}[t]{@{}r@{\hspace{0.8em}} cr l@{}}
  \mbox{\emph{Goals:}}
& G & ::= & a
      \sep  (\bigLam{C}) \Imp G
      \sep  \A{x}{G}
\\ \mbox{\emph{Clauses:}}
& C & ::= & \makebox[3em]{$\Eq{a}{\alpha}$}
      \sep  C \And G
      \sep  \E{x}{C}
\end{array}$
\hfill
$\begin{array}[t]{@{}r@{\hspace{0.8em}} cr l@{}}
 \mbox{\emph{Programs:}}
& \Psi & ::= & \cdot
       \sep  \Psi, \bigLam{C}
\end{array}$
\hspace*{1em}}%

\medskip

The operational semantics of a compiled program, as given by the above
grammar, is defined on the basis of the following two judgments:
$$
\begin{array}{l@{\hspace{1.5em}}p{16em}}
   \gResA{\Psi}{G}
 & \emph{$G$ is uniformly provable from $\Psi$}
\\ \rResA{\Psi}{\C}
 & \emph{$\C$ is uniformly provable from $\Psi$}
\end{array}
$$
Here, clause instances $\C$ are $C$'s whose variable $\alpha$ has been
instantiated with an atomic formula $a'$.
%
%
%
The operational semantics of $\Lc$ is shown in
Figure~\ref{fig:pil0-resolution}.  Observe that, with the partial exception of
\rname{g0\_atm}, it consists solely of right rules.  This means that every
connective is seen as a search directive: the dynamic clause preparations
embodied by the left rules has now been turned into right search rules through
a static compilation phase.

\begin{figure}[t]
\begin{center}
\leavevmode
\newcommand{\rsp}{1.8ex}
\newcommand{\rspl}{-.5ex}
\newcommand{\trule}[1][-1.5ex]{\rule[#1]{0em}{0ex}}
\newcommand{\hhfill}{\hfill}
\fboxsep=0pt

\noindent
\fbox{\scriptsize
$\begin{array}{@{\hfill}c@{\hfill}}
\makebox[\textwidth]{}\\[-2.5ex]
\multicolumn{1}{l}{\bf \scriptstyle Goals\trule}
\\
  \cRule
   {\rResA{\Psi, \bigLam{C}, \Psi'}{[a/\alpha]C}}
   {\gResA{\Psi, \bigLam{C}, \Psi'}{a}}
   {g0\_atm}
\hhfill
  \cRule
   {\gResA{\Psi, \bigLam{C}}{G}}
   {\gResA{\Psi}{(\bigLam{C}) \Imp G}}
   {g0\_imp}
\hhfill
  \cRule
   {c \; \mbox{\em ``new''}
    \qq
    \gResA{\Psi}{[c/x]G}}
   {\gResA{\Psi}{\A{x}{G}}}
   {g0\_all}
\\[\rspl]
\hline\\[-4.5ex]
\multicolumn{1}{l}{\bf \scriptstyle Clause\:instances\trule}
\\[-1.0ex]
  \cRule
   {}
   {\rResA{\Psi}{\Eq{a}{a}}}
   {r0\_eq}
\hhfill
  \cRule
   {\rResA{\Psi}{\C}
    \qq
    \gResA{\Psi}{G}}
   {\rResA{\Psi}{\C \And G}}
   {r0\_and}
\hhfill
  \cRule
   {\rResA{\Psi}{[t/x]\C}}
   {\rResA{\Psi}{\E{x}{\C}}}
   {r0\_exists}
\\[-1ex]\relax
\end{array}$}

\caption{Search Semantics of $\Lc$.}
\label{fig:pil0-resolution}
\end{center}
\end{figure}

\subsection{Compilation}
\label{sec:pil0-compilation}

Compilation, the process that transforms a logic program in $\Ls$ into a
compiled program in $\Lc$, is expressed by means of the following three
judgments:\pagebreak[3]
$$
\begin{array}{l@{\hspace{1.5em}}p{16em}}
   \cmpProg{\G}{\Psi}
 & \emph{Program $\G$ is compiled to $\Psi$}
\\ \cmpClause{A}{C}
 & \emph{Clause $A$ with $\alpha$ is compiled to $C$ }
\\ \cmpGoal{A}{G}
 & \emph{Goal $A$ is compiled to $G$}
\end{array}
$$
These judgments are defined by the rules in
Figure~\ref{fig:pil0-compilation} --- see~\cite{Cervesato98jicslp} for details.

As our ongoing example, consider the following two clauses, taken from a type
checking specification for a Church-style simply typed $\lambda$-calculus.
For clarity, we write program clauses Prolog-style, using the reverse
implication $\Pim$ instead of $\Imp$ in positive formulas.
\begin{enumerate}\small
\item%
$\begin{array}[t]{@{}l@{\hspace{3.5em}}c@{\hspace{2em}}l@{}}
  \begin{array}[t]{@{}ll@{}}
  \\ \multicolumn{2}{@{}l@{}}{
       \A{E_1}{}\A{E_2}{}
       \A{T_1}{}\A{T_2}{}}
  \\      & \ms{of}\;(\ms{app}\;E_1\;E_2)\;T_2
  \\ \Pim & \ms{of}\;E_1\;(\ms{arr}\;T_1\;T_2)
  \\ \Pim & \ms{of}\;E_2\;T_1
  \end{array}
& \raisebox{-7ex}{ \ $\cmpProg{}{}$ \ }
& \begin{array}[t]{@{}ll@{}}
     \multicolumn{2}{@{}l@{}}{\bigLam{}}
  \\ \multicolumn{2}{@{}l@{}}{
       \E{E_1}{}\E{E_2}{}
       \E{T_1}{}\E{T_2}{}}
  \\      & \Eq{(\ms{of}\;(\ms{app}\;E_1\;E_2)\;T_2)}{\alpha}
  \\ \And & \ms{of}\;E_1\;(\ms{arr}\;T_1\;T_2)
  \\ \And & \ms{of}\;E_2\;T_1
  \end{array}
\end{array}$


\item%
$\begin{array}[t]{@{}lcl@{}}
  \begin{array}[t]{@{}ll@{}}
  \\ \multicolumn{2}{@{}l@{}}{
       \A{E}{}
       \A{T_1}{}\A{T_2}{}}
  \\      & \ms{of}\;(\ms{lam}\;T_1\;E)\;(\ms{arr}\;T_1\;T_2)
  \\ \Pim & (\A{x}{} \ms{of}\;x\;T_1
  \\      & \hspace{0.5em} \Imp \ms{of}\;(E\;x)\;T_2)
  \end{array}
& \raisebox{-6ex}{$\cmpProg{}{}$}
& \begin{array}[t]{@{}ll@{}}
     \multicolumn{2}{@{}l@{}}{\bigLam{}}
  \\ \multicolumn{2}{@{}l@{}}{
       \E{E}{}
       \E{T_1}{}\E{T_2}{}}
  \\      & \Eq{(\ms{of}\;(\ms{lam}\;T_1\;E)\;(\ms{arr}\;T_1\;T_2))}{\alpha}
  \\ \And & (\A{x}{} \begin{array}[t]{@{}l@{\;}l@{}}
                             & \bigLam[\beta]{(\Eq{(\ms{of}\;x\;T_1)}{\beta})}
                     \\ \Imp & \ms{of}\;(E\;x)\;T_2)
                     \end{array}
  \end{array}
\end{array}$

\end{enumerate}

The compiled language $\Lc$ is sound and complete for $\Ls$.
See~\cite{Cervesato98jicslp} for the formal statements.  The proof of both
directions proceeds by straightforward induction, which contrasts greatly with
the complex proofs of soundness and correctness previously devised for the
WAM~\cite{Boerger95fmpa,Russinoff92jlp}.



\begin{figure}[t]
\begin{center}
\leavevmode
\newcommand{\rsp}{1.8ex}
\newcommand{\rspl}{-.5ex}
\newcommand{\trule}[1][-1.5ex]{\rule[#1]{0em}{0ex}}
\newcommand{\hhfill}{\hfill}
\fboxsep=0pt

\noindent
\fbox{\scriptsize
$\begin{array}{@{\hfill}c@{\hfill}}
\makebox[\textwidth]{}\\[-2.5ex]
\multicolumn{1}{l}{\bf \scriptstyle Programs\trule}
\\[-0.5ex]
  \cRule
   {}
   {\cmpProg{\cdot}{\cdot}}
   {p0c\_empty}
\hhfill
  \cRule
   {\cmpProg{\G}{\Psi}
    \qq
    \cmpClause{A}{C}}
   {\cmpProg{\G, A}{\Psi, \bigLam{C}}}
   {p0c\_clause}
\\[\rspl]
\hline\\[-4.5ex]
\multicolumn{1}{l}{\bf \scriptstyle Clauses\trule}
\\[-1.0ex]
  \cRule
   {}
   {\cmpClause{a}{\Eq{a}{\alpha}}}
   {c0c\_atm}
\hhfill
  \cRule
   {\cmpClause{B}{C}
    \qq
    \cmpGoal{A}{G}}
   {\cmpClause{A \Imp B}{C \And G}}
   {c0c\_imp}
\hhfill
  \cRule
   {\cmpClause{A}{C}}
   {\cmpClause{\A{x}{A}}{\E{x}{C}}}
   {c0c\_all}
\\[\rspl]
\hline\\[-4.5ex]
\multicolumn{1}{l}{\bf \scriptstyle Goals\trule}
\\[-1.0ex]
  \cRule
   {}
   {\cmpGoal{a}{a}}
   {g0c\_atm}
\hhfill
  \cRule
   {\cmpClause{A}{C}
    \qq
    \cmpGoal{B}{G}}
   {\cmpGoal{A \Imp B}{(\bigLam{C}) \Imp G}}
   {g0c\_imp}
\hhfill
  \cRule
   {\cmpGoal{A}{C}}
   {\cmpGoal{\A{x}{A}}{\A{x}{C}}}
   {g0c\_all}
\\[-1ex]\relax
\end{array}$}

\caption{Compilation of $\Ls$ into $\Lc$.}
\label{fig:pil0-compilation}
\end{center}
\end{figure}

\section{Fully Logical Compilation}
\label{sec:pil1} 

Because clauses are compiled to expressions of the form $\bigLam{C}$, the
language $\Lc$ is not fully logical.  In this section we consider a different
compilation target, the language $\LcB$, which lies entirely within logic.


In the previous section, a generic Horn clause of the form
\begin{equation}
\label{eq:horn}
\A{\vec{y}}{(p\:\vec{t} \Pim a_1 \Pim \ldots \Pim a_n)}
\end{equation}
was compiled into
$\bigLam{\E{\vec{y}}{(\Eq{p\:\vec{t}}{\alpha} \And a_1 \And \ldots \And a_n)}}.$
During execution, rule \rname{c0\_atm} reduced the current atomic goal $a$ to
the clause instance $\E{\vec{y}}{(\Eq{p\:\vec{t}}{a} \And a_1 \And \ldots
  \And a_n)}$.  Note that $\vec{t}$ may depend on $\vec{y}$, but $a$ does not.
We will now compile that Horn clause into
\begin{equation}
\label{eq:pil1-horn}
\A{\vec{x}}{(
        p\;\vec{x}
  \Pim \E{\vec{y}}{(\Eq{\vec{x}}{\vec{t}} \And a_1 \And \ldots \And a_n}))}
\end{equation}%
where $\vec{x}$ is a sequence of fresh variables, all distinct from each
other, and equal in number to the arity of $p$, and $\Eq{\vec{x}}{\vec{t}}$
stands for a conjunction of equalities between each variable $x_i$ in
$\vec{x}$ and the term $t_i$ in $\vec{t}$ in the corresponding position (or
$\True$ if the arity of $p$ is zero).  Notice that the non-logical
second-order binder ``$\bigLam{\!}$'' is gone.  At run time,
formula~(\ref{eq:pil1-horn}) will resolve an atomic goal $p\;\vec{t'}$ into
the clause $p\;\vec{t'} \Pim \E{\vec{y}}{(\Eq{\vec{t'}}{\vec{t}} \And a_1 \And
  \ldots \And a_n)}$, which immediately reduces to
$\E{\vec{y}}{(\Eq{\vec{t'}}{\vec{t}} \And a_1 \And \ldots \And a_n})$.  Like
earlier, $\vec{t}$ may depend on $\vec{y}$, but $\vec{t'}$ does not.  The
variables $\vec{x}$ correspond directly to the ``argument registers''
(\verb"A"$n$) of the WAM~\cite{Aitkaci91book}, while the $\vec{y}$'s are closely
related to its ``permanent variables'' (\verb"Y"$n$).

Formula~(\ref{eq:pil1-horn}) can be understood as an uncurried form
of~(\ref{eq:horn}): outer implications are transformed into conjunctions and
universals into existentials.  Doing so literally would yield the formula
$p\;\vec{t} \Pim \E{\vec{y}}{(a_1 \And \ldots \And a_n)}$, which is incorrect
because occurrences of variables in $\vec{y}$ within $\vec{t}$ have escaped
their scope.  Instead, formula~(\ref{eq:pil1-horn}) installs fresh variables
$\vec{x}$ as the arguments of the head predicate $p$ and adds the equality
constraints $\Eq{\vec{x}}{\vec{t}}$ in the body.

\subsection{Target Language}
\label{sec:pil1-target}

We now generalize the above intuition to any formula in $\Ls$, not just Horn
clauses.  Our second target language, $\LcB$, is given by the following
grammar. 

\medskip
\noindent
{\arraycolsep=0.25em
\newcommand{\sep}{\hspace{0.5em}|\hspace{0.5em}}
$\begin{array}[t]{@{}r@{\hspace{0.8em}} cr l@{}}
  \mbox{\emph{Goals:}}
& G & ::= & a
      \sep  C \Imp G
      \sep  \A{x}{G}
\\ \mbox{\emph{Clauses:}}
& C & ::= & R \Imp p\:\vec{x}
      \sep  \A{x}{C}
\\ \mbox{\emph{Residuals:}}
& R & ::= & \Eq{x}{t}
      \sep  \True
      \sep  R \And G
      \sep  \E{x}{R}
\end{array}$
\hfill
$\begin{array}[t]{@{}r@{\hspace{0.8em}} cr l@{}}
 \mbox{\emph{Programs:}}
& \Psi & ::= & \cdot
       \sep  \Psi, C
\end{array}$
\hspace*{1em}}%

\medskip
\noindent
Compiled goals ($G$) are just like in Section~\ref{sec:pil0-target}: atoms,
hypothetical goals, or universal goals.  Compiled clauses ($C$) have the form
$\A{\vec{x}}{(R \Imp p\:\vec{x})}$, i.e., a (possibly empty) outer layer of
universal quantifiers enclosing an implication $R \Imp p\:\vec{x}$ whose head
$p\:\vec{x}$ always consists of a predicate name ($p$) applied to a (possibly
empty) sequence of distinct variables ($\vec{x}$).  Its body is a
\emph{residual} ($R$).  A residual can be either an equality constraint
($\Eq{x}{t}$), the trivial constraint $\True$ (logical truth), or like in
Section~\ref{sec:pil0-target} a goal invocation or an instantiation request.
Notice that $C$ is now the full result of compiling a clause.

\begin{figure}[t]
\begin{center}
\leavevmode
\newcommand{\rsp}{1.8ex}
\newcommand{\rspl}{-.5ex}
\newcommand{\trule}[1][-1.5ex]{\rule[#1]{0em}{0ex}}
\newcommand{\hhfill}{\hfill}
\fboxsep=0pt

\noindent
\fbox{\scriptsize
$\begin{array}{@{\hfill}c@{\hfill}}
\makebox[\textwidth]{}\\[-2.5ex]
\multicolumn{1}{l}{\bf \scriptstyle Goals\trule}
\\[-0.5ex]
  \cRule
   {\cResB{\Psi, C, \Psi'}{C}}
   {\gResB{\Psi, C, \Psi'}{a}}
   {g1\_atm}
\hhfill
  \cRule
   {\gResB{\Psi, C}{G}}
   {\gResB{\Psi}{C \Imp G}}
   {g1\_imp}
\hhfill
  \cRule
   {c \; \mbox{\em ``new''}
    \qq
    \gResB{\Psi}{[c/x]G}}
   {\gResB{\Psi}{\A{x}{G}}}
   {g1\_all}
\\[\rspl]
\hline\\[-4.5ex]
\multicolumn{1}{l}{\bf \scriptstyle Clauses\trule}
\\[-1.0ex]
  \cRule
   {\rResB{\Psi}{\R}}
   {\cResB{\Psi}{\R \Imp a}}
   {c1\_imp}
\hhfill
  \cRule
   {\cResB{\Psi}{[t/x]\C}}
   {\cResB{\Psi}{\A{x}{\C}}}
   {c1\_all}
\\[\rspl]
\hline\\[-4.5ex]
\multicolumn{1}{l}{\bf \scriptstyle Residuals\trule}
\\[-1.0ex]
  \cRule
   {}
   {\gResB{\Psi}{\Eq{t}{t}}}
   {r1\_eq}
\hhfill
  \cRule
   {}
   {\rResB{\Psi}{\True}}
   {r1\_true}
\hhfill
  \cRule
   {\rResB{\Psi}{\R}
    \qq
    \gResB{\Psi}{G}}
   {\rResB{\Psi}{\R \And G}}
   {r1\_and}
\hhfill
  \cRule
   {\rResB{\Psi}{[t/x]\R}}
   {\rResB{\Psi}{\E{x}{\R}}}
   {r1\_exists}
\\[-1ex]\relax
\end{array}$}

\caption{Search Semantics of $\LcB$.}
\label{fig:pil1-resolution}
\end{center}
\end{figure}

The operational semantics of $\LcB$ is specified by the following three
judgments:
$$
\begin{array}{l@{\hspace{1.5em}}p{18em}}
   \gResB{\Psi}{G}
 & \emph{$G$ is uniformly provable from $\Psi$}
\\ \cResB{\Psi}{\C}
 & \emph{$a$ is immediately entailed by $\C$ in $\Psi$}
\\ \rResB{\Psi}{\R}
 & \emph{$\R$ is uniformly provable from $\Psi$}
\end{array}
$$
where $\C$ and $\R$ differ from $C$ and $R$ by the instantiation of some
variables in a clause head and on the left-hand side of equalities, respectively.


Their operational semantics is given in Figure~\ref{fig:pil1-resolution}.
Goals are handled exactly in the same way as uniform provability in $\Ls$ (top
part of Figure~\ref{fig:pil0-uniform}).  The operational reading of compiled
clauses is an instance of that of immediate entailment: rule \rname{c1\_imp} is
a special case of \rname{i\_imp} while \rname{c1\_all} is isomorphic to
\rname{i\_all}.  Note that rule \rname{c1\_imp} reduces immediately to the
residual $R$ if the head of the clause matches the atomic goal $a$ being
proved.  The rules for residuals correspond closely to the rules for clause
instances for our original target language at the bottom of
Figure~\ref{fig:pil0-resolution}: rule \rname{r1\_eq} requires that the two
sides of an equality be indeed equal and rule \rname{r1\_true} is always
satisfied.


The rules in Figure~\ref{fig:pil1-resolution} build uniform
proofs~\cite{Miller91apal}, characteristic of abstract logic programming
languages: the operational semantics decomposes a goal to an atomic formula
(top segment of Figure~\ref{fig:pil1-resolution}), then selects a clause and
focuses on it until it finds a matching head (middle segment) and then
decomposes its body (bottom segment), which may eventually expose some goals, and
the cycle repeats.  In particular, once an atomic goal $p\:\vec{t}$ has been
exposed, a successful derivation will necessarily contain an instance of rule
\rname{g1\_atm} that picks a clause $C$ with head $p\:\vec{x}$, as many
instances of rule \rname{c1\_all} as the arity of $p$, and an instance of rule
\rname{c1\_imp}.  This necessary sequence of steps is captured by the following
derived ``macro-rule'' (the \emph{backchaining} rule):
$$
\cRule
 {\rResB{\Psi, \A{\vec{x}}{(R \Imp p\:\vec{x})}, \Psi'}{[\vec{t}/\vec{x}]R}}
 {\gResB{\Psi, \A{\vec{x}}{(R \Imp p\:\vec{x})}, \Psi'}{p\:\vec{t}}}
 {g1\_atm'}
$$
Replacing rules \rname{g1\_atm}, \rname{c1\_all} and \rname{c1\_imp} with rule
\rname{g1\_atm'} 
yields a system that is equivalent to that in
Figure~\ref{fig:pil1-resolution}.  Taking it as primitive amounts to replacing
the construction for compiled clauses, $\A{\vec{x}}{(R \Imp p\:\vec{x})}$,
with a synthetic connective, call it $\Lam_p \vec{x}.\, R$.  Therefore, by
accounting for the structure of atomic propositions and proper quantification
patterns, $\LcB$ provides a fully logical justification for clause compilation
that $\Lc$'s $\Lam \alpha.\, C$ lacked.




\subsection{Compilation}
\label{sec:pil1-compilation}

Compilation transforms logic programs in $\Ls$ into compiled logic programs in
$\LcB$.  In order to define it, the auxiliary notion of pseudo clause will
come handy:

\medskip
\noindent
{\arraycolsep=0.25em
\newcommand{\sep}{\hspace{0.5em}|\hspace{0.5em}}
$\begin{array}{@{}r@{\hspace{0.8em}} cr l@{}}
  \mbox{\emph{Pseudo Clauses:}}
& \CC & ::= & \Box \Imp p\:\vec{x}
      \sep  \A{x}{\CC}
\end{array}$}%
\medskip

\noindent
A pseudo clause retains the outer structure of a clause, but has a hole
($\Box$) in place of the residual $R$.  In general, a pseudo clause $\CC$ has
the form $\A{\vec{x}}{\Box \Imp p\:\vec{x'}}$.  In a fully compiled clause,
variables $\vec{x}$ will coincide with $\vec{x'}$.

Pseudo clauses are generated while processing the head of a clause.  The hole
then needs to be replaced with the compiled body, a residual.  We write this
operation, pseudo clause instantiation, as $\CC[R]$.  It is formally defined
as follows:
$$
\left\{
\begin{array}{lcl}
   (\Box \Imp p\:\vec{x})[R] & = & R \Imp p\:\vec{x}
\\ (\A{x}{\CC})[R]   & = & \A{x}{(\CC[R])}
\end{array}
\right.
$$
As is often the case with such contextual operations, pseudo clause
instantiation can, and generally will, lead to variable capture: in
$(\A{\vec{x}}{\Box \Imp p\:\vec{x}})[R]$, there may be free occurrences of
variables in $\vec{x}$ within $R$.  In the result, these occurrences are
bound by the outer quantifiers.

Compilation is expressed by means of the following four judgments
$$
\begin{array}{l@{\hspace{1.5em}}p{17em}}
   \cmpProgB{\G}{\Psi}
 & \emph{Program $\G$ is compiled to $\Psi$}
\\ \cmpHeadB{\vec{x}}{a}{\CC}{E}
 & \emph{Head $a$ with $\vec{x}$ is compiled to $\CC$ and $E$}
\\ \cmpClauseB{A}{R}
 & \emph{Clause $A$ is compiled to $\CC$ and $R$}
\\ \cmpGoalB{A}{G}
 & \emph{Goal $A$ is compiled to $G$}
\end{array}
$$
and defined by the rules in Figure~\ref{fig:pil1-compilation}, where we wrote
$E$ for conjunctions of equalities.  The judgment $\cmpClauseB{A}{R}$ compiles
an $\Ls$ clause $A$ into a pseudo clause $\CC$ and a residual $R$.  They are
assembled into an $\LcB$ clause in rules \rname{p1c\_clause} and
\rname{g1c\_imp}.  Programs and goals are otherwise compiled just as for
$\Lc$ in Figure~\ref{fig:pil0-compilation}.  Clause heads are handled
differently: rule \rname{c1c\_atm} invokes the auxiliary head compilation
judgment to compile the goal $p\:\vec{t}$ into a pseudo clause
$\A{\vec{x}}{\Box \Imp p\:\vec{x}}$ and the equalities
$\Eq{\vec{x}}{\vec{t}}$, which will form the seed of the clause's residual.

Consider the first example clause in Section~\ref{sec:pil0-compilation}.  Its
head ($\ms{of}\;(\ms{app}\;E_1\;E_2)\;T_2$) is compiled into the pseudo clause
$\A{x_1}{\A{x_2}{(\Box \Imp \ms{of}\;x_1\;x_2)}}$ and the equality constraints
$\True \And (\Eq{x_1}{\ms{app}\;E_1\;E_2}) \And (\Eq{x_2}{T_2})$, where $x_1$
and $x_2$ are new variables.  These core equalities are then extended with the
compiled body of that clause, $(\ms{of}\;E_1\;(\ms{arr}\;T_1\;T_2)) \And
(\ms{of}\;E_2\;T_1)$, and existential quantifications over the original
variables of the clause, $E_1$, $E_2$, $T_1$ and $T_2$, are finally wrapped
around the result before embedding it in the hole of the pseudo clause.  The
resulting $\LcB$ clause is displayed in the top part of
Figure~\ref{fig:pl1-example}.

\begin{figure}[t]
\begin{center}
\leavevmode
\newcommand{\rsp}{1.8ex}
\newcommand{\rspl}{-.5ex}
\newcommand{\trule}[1][-1.5ex]{\rule[#1]{0em}{0ex}}
\newcommand{\hhfill}{\hfill}
\fboxsep=0pt

\noindent
\fbox{\scriptsize
$\begin{array}{@{\hfill}c@{\hfill}}
\makebox[\textwidth]{}\\[-3ex]
\multicolumn{1}{l}{\bf \scriptstyle Programs\trule}
\\[-0.5ex]
  \cRule
   {}
   {\cmpProgB{\cdot}{\cdot}}
   {p1c\_empty}
\hhfill
  \cRule
   {\cmpProgB{\G}{\Psi}
    \qq
    \cmpClauseB[\CC]{A}{R}}
   {\cmpProgB{\G, A}{\Psi, \CC[R]}}
   {p1c\_clause}
\\[\rspl]
\hline\\[-4.5ex]
\multicolumn{1}{l}{\bf \scriptstyle Heads\trule}
\\[-1.0ex]
  \cRule
   {}
   {\cmpHeadB{\vec{x}}{p}{\Box \Imp p\:\vec{x}}{\True}}
   {h1c\_p}
\hhfill
  \cRule
   {\cmpHeadB{x\:\vec{x}}{a}{\CC}{E}
    \qq
    x \: \mbox{\em ``new''}}
   {\cmpHeadB{\vec{x}}{a\;t}{\A{x}{\CC}}{E \And \Eq{x}{t}}}
   {h1c\_pt}
\\[\rspl]
\hline\\[-4.5ex]
\multicolumn{1}{l}{\bf \scriptstyle Clauses\trule}
\\[0.0ex]
  \cRule
   {\cmpHeadB{\cdot}{a}{\CC}{E}}
   {\cmpClauseB[\CC]{a}{E}}
   {c1c\_atm}
\hhfill
  \cRule
   {\cmpGoalB{A}{G}
    \qq
    \cmpClauseB[\CC]{B}{R}}
   {\cmpClauseB[\CC]{A \Imp B}{R \And G}}
   {c1c\_imp}
\hhfill
  \cRule
   {\cmpClauseB[\CC]{A}{R}}
   {\cmpClauseB[\CC]{\A{x}{A}}{\E{x}{R}}}
   {c1c\_all}
\\[\rspl]
\hline\\[-4.5ex]
\multicolumn{1}{l}{\bf \scriptstyle Goals\trule}
\\[-1.0ex]
  \cRule
   {}
   {\cmpGoalB{a}{a}}
   {g1c\_atm}
\hhfill
  \cRule
   {\cmpClauseB[\CC]{A}{R}
    \qq
    \cmpGoalB{B}{G}}
   {\cmpGoalB{A \Imp B}{\CC[R] \Imp G}}
   {g1c\_imp}
\hhfill
  \cRule
   {\cmpGoalB{A}{C}}
   {\cmpGoalB{\A{x}{A}}{\A{x}{C}}}
   {g1c\_all}
\\[-1ex]\relax
\end{array}$}

\caption{Compilation of $\Ls$ into $\LcB$.}
\label{fig:pil1-compilation}
\end{center}
\end{figure}

The target language $\LcB$ is sound and complete with respect to $\Ls$.  In
order to show it, we need the following auxiliary results.  The first
statement is proved by induction on the structure of $a$.  The second by
induction on the given derivation.

\noindent{\parbox{\linewidth}{%
\begin{lemma}
\begin{itemize}
\itemsep=0pt
\item%
  If $\cmpHeadB{\vec{x}}{a}{\CC}{E}$, then for all $\vec{t}$ of
  the same length as $\vec{x}$ and all $\Psi$ we
  have $\cResB[a\:\vec{t}]{\Psi}{[\vec{t}/\vec{x}](\CC[E])}$.
\item%
  If $\cResB{\Psi}{\CC[R]}$, then $\rResB{\Psi}{R}$.
\end{itemize}
\end{lemma}}

The statements of soundness and completeness are as follows.  For each of
them, the proof proceeds by mutual induction on the first derivation in the
antecedent.

\noindent{\parbox{\linewidth}{%
\begin{theorem}[Soundness of the compilation to $\LcB$]
\label{th:pil1-sound}
\begin{itemize}
\itemsep=0ex
\item%
    If   \ $\unid{\G}{A}$,
         \ $\cmpProgB{\G}{\Psi}$
  \ and  \ $\cmpGoalB{A}{G}$,
  \ then \ $\gResB{\Psi}{G}$.
\item%
    If   \ $\immd{\G}{A}$,
         \ $\cmpProgB{\G}{\Psi}$
  \ and  \ $\cmpClauseB[\CC]{A}{R}$,
  \ then \ $\cResB{\Psi}{\CC[R]}$.
\end{itemize}
\end{theorem}}

\noindent{\parbox{\linewidth}{%
\begin{theorem}[Completeness of the compilation to $\LcB$]
\label{th:pil1-complete}
\begin{itemize}
\itemsep=0ex
\item%
    If   \ $\gResB{\Psi}{G}$,
         \ $\cmpProgB{\G}{\Psi}$
  \ and  \ $\cmpGoalB{A}{G}$,
  \ then \ $\unid{\G}{A}$.
\item%
    If   \ $\cResB{\Psi}{C}$,
         \ $\cmpProgB{\G}{\Psi}$,
         \ $C = \CC[R]$
  \ and  \ $\cmpClauseB[\CC]{A}{R}$,
  \ then \ $\immd{\G}{A}$.
\end{itemize}
\end{theorem}}

\begin{figure}[t]
\fbox{%
\parbox{\linewidth}{%
\begin{enumerate}\small
\item%
$\begin{array}[t]{@{}l@{\hspace{3.5em}}c@{\hspace{2em}}l@{}}
  \begin{array}[t]{@{}ll@{}}
     \multicolumn{2}{@{}l@{}}{
       \A{E_1}{}\A{E_2}{}
       \A{T_1}{}\A{T_2}{}}
  \\      & \ms{of}\;(\ms{app}\;E_1\;E_2)\;T_2
  \\
  \\
  \\
  \\ \Pim & \ms{of}\;E_1\;(\ms{arr}\;T_1\;T_2)
  \\ \Pim & \ms{of}\;E_2\;T_1
  \end{array}
& \raisebox{-10ex}{ \ $\cmpProgB{}{}$ \ }
& \begin{array}[t]{@{}ll@{}l@{\;}l@{}}
     \multicolumn{4}{@{}l@{}}{\A{x_1}{}\A{x_2}{}}
  \\      & \multicolumn{3}{@{}l@{}}{\ms{of}\;x_1\;x_2}
  \\ \Pim &(&\multicolumn{2}{@{}l@{}}{
          \E{E_1}{}\E{E_2}{}
          \E{T_1}{}\E{T_2}{\True}}
  \\ && \And & \Eq{x_1}{\ms{app}\;E_1\;E_2}
  \\ && \And & \Eq{x_2}{T_2}
  \\ && \And & \ms{of}\;E_1\;(\ms{arr}\;T_1\;T_2)
  \\ && \And & \ms{of}\;E_2\;T_1)
  \end{array}
\end{array}$


\medskip
\item%
$\begin{array}[t]{@{}lcl@{}}
  \begin{array}[t]{@{}ll@{}}
     \multicolumn{2}{@{}l@{}}{
       \A{E}{}
       \A{T_1}{}\A{T_2}{}}
  \\      & \ms{of}\;(\ms{lam}\;T_1\;E)\;(\ms{arr}\;T_1\;T_2)
  \\
  \\
  \\
  \\ \Pim & (\A{x}{} 
  \\
  \\
  \\      & \hspace{2em} \ms{of}\;x\;T_1
  \\      & \hspace{0.5em} \Imp \ms{of}\;(E\;x)\;T_2)
  \end{array}
& \raisebox{-14.5ex}{$\cmpProgB{}{}$}
& \begin{array}[t]{@{}ll@{}l@{\;}l@{}}
     \multicolumn{4}{@{}l@{}}{\A{x_1}{}\A{x_2}{}}
  \\      & \multicolumn{3}{@{}l@{}}{\ms{of}\;x_1\;x_2}
  \\ \Pim &(&\multicolumn{2}{@{}l@{}}{
          \E{E}{}
          \E{T_1}{}\E{T_2}{\True}}
  \\ &&\And & \Eq{x_1}{\ms{lam}\;T_1\;E}
  \\ &&\And & \Eq{x_2}{\ms{arr}\;T_1\;T_2}
  \\ &&\And & (\A{x}{} \begin{array}[t]{@{}l@{\;}l@{}}
                             & \begin{array}[t]{@{}l@{\;}l@{}}
                                  \multicolumn{2}{@{}l@{}}{\A{x_1'}{}\A{x_2'}{\True}}
                               \\ \And & \Eq{x_1'}{x}
                               \\ \And & \Eq{x_2'}{T_1}
                               \\ \And & \ms{of}\;x_1'\;x_2')
                               \end{array}
                     \\ \Imp & \ms{of}\;(E\;x)\;T_2)
                     \end{array}
  \end{array}
\end{array}$
\end{enumerate}}}
\label{fig:pl1-example}
\caption{$\LcB$ Compilation Example}
\end{figure}

We conclude this section by showing in Figure~\ref{fig:pl1-example} the output
of our compilation procedure for the two examples seen in
Section~\ref{sec:pil0-compilation}.  We stretch the source clauses (left) to
align corresponding atoms.  As can be gleaned from these clauses, there are
ample opportunities for optimizations in our compilation process.  In
particular, a constraint $\Eq{x}{y}$ mentioning variables on both sides can
often be eliminated by replacing the existential variable $y$ with the
universal variable $x$ in the rest of the clause (and removing the existential
quantifier) --- the exception is when there are multiple constraints of this
form for the same $y$.  The leading logical constant $\True$ makes for a
succinct presentation of the compilation process, but plays no actual role: it
can also be eliminated.

It is interesting to rewrite these clauses using the synthetic connective
$\Lam_p$ discussed earlier (we have omitted occurrences of $\True$ for readability):
$$
\begin{array}{lll}
    \Lam_{\ms{of}}\; x_1\:x_2.
  & \multicolumn{2}{l}{\E{E_1}{}\E{E_2}{}\E{T_1}{}\E{T_2}{}}
\\&       & \Eq{x_1}{\ms{app}\;E_1\;E_2}
     \And\; \Eq{x_2}{T_2}
\\&  \And & \ms{of}\;E_1\;(\ms{arr}\;T_1\;T_2)
   \;\And\; \ms{of}\;E_2\;T_1
\\[1ex]
    \Lam_{\ms{of}}\; x_1\:x_2.
  & \multicolumn{2}{l}{\E{E}{}\E{T_1}{}\E{T_2}{}}
\\&       & \Eq{x_1}{\ms{lam}\;T_1\;E}
   \;\And\; \Eq{x_2}{\ms{arr}\;T_1\;T_2}
\\&  \And & \A{x}{}(\Lam_{\ms{of}}\; x_1'\:x_2'.\;\;
                                 \Eq{x_1'}{x}  \:\And\:
                                 \Eq{x_2'}{T_1})
                     \;\Imp\; \ms{of}\;(E\;x)\;T_2
\end{array}
$$

\section{Support for Moded Programs}
\label{sec:pil2}

In this section, we will specialize the compilation process just outlined to
the case where the source program is well-moded.  In a well-model program, the
argument positions of each predicate symbol are designated as either input or
output.  Input arguments are guaranteed to be ground terms at the time a goal
is called.  Dually, output arguments are guaranteed to have been made ground
by the time the call returns.

There are operational benefits to working with well-moded programs: while an
interpreter for a generic program must implement term-level unification,
well-moded programs can be executed by relying uniquely on pattern matching
and variable instantiation.  This is desirable because matching often behaves
better than general unification.  For example, it is more efficient for
first-order term languages were it only because it does away with the
occurs-check, and it is decidable for higher-order term languages while
general unification is not~\cite{stirling09lmcs}.

The development in this section is motivated by well-moding, but is sound
independently of whether a program is well-moded or not.  Statically enforcing
well-moding brings the operational advantages just discussed, but the results
in this section do not depend on it.

\subsection{Source Language}
\label{sec:pil2-source}

In this section, we assume that each predicate symbol in $\Ls$ comes with a
\emph{mode} which declares each of its arguments as input, written $\IN{\;}$,
or output, written $\OT{\;}$.  For simplicity of exposition, we decorate the
actual arguments of all atomic propositions with these symbols, so that a term
$t$ in input position in an atomic proposition is written $\IN{t}$ (read ``in
$t$'').  Similarly $t$ in output position is written $\OT{t}$ (pronounced
``out $t$'').  This amounts to revising the grammar of atomic propositions as
follows:


\medskip
\noindent
{\arraycolsep=0.25em
\newcommand{\sep}{\hspace{0.5em}|\hspace{0.5em}}
$\begin{array}{@{}r@{\hspace{0.8em}} cr l@{}}
  \mbox{\emph{Atoms:}}
& a & ::= & p
      \sep a\:\IN{t}
      \sep a\:\OT{t}
\end{array}$}

\medskip
\noindent
Just like we assume that the arity of a predicate symbol $p$ remains constant
in a program, we require that all atomic propositions for $p$ have their
input/output marks in the same positions.  This pattern is the mode of $p$ ---
an actual language would rely on explicit mode declarations.

For typographic convenience and without loss of generality, our examples
assume that input positions precede output positions so that an atomic formula
$a$ can be written as $p\;\Ivec{t}\;\Ovec{t}$ where $\Ivec{t}$ and $\Ovec{t}$
are the (possibly empty) sequences of terms in input (resp.\ output) positions
for $p$.  To avoid notational proliferation, we use the markers $\IN{\;}$ and
$\OT{\;}$ both as mode designators and as symbol decorations (like primes and
subscripts) when working with generic terms.  Therefore, $\IN{t}$ and $\OT{t}$
indicate possibly different terms in \mbox{$p\:\IN{t}\:\OT{t}$}, and similarly
for term sequences, as in $p\;\Ivec{t}\;\Ovec{t}$ above.


At our level of abstraction, the rules in Figure~\ref{fig:pil0-uniform}
capture the operational semantics of this variant of $\Ls$: mode annotations
are simply ignored.  However, moded execution requires that two of the
operational choices left open by those rules be resolved using some
algorithmic strategy: the order in which rule \rname{i\_imp} searches for
derivations of its two premises, and the substitution term that rule
\rname{i\_all} picks.  For both, we will assume the same strategy as Prolog:
implement rule \rname{i\_imp} left to right and implement rule \rname{i\_all}
lazily by replacing each variable $x$ with a ``logical variable'' $X$ which is
instantiated incrementally through unification.  This allows us to view an
atomic goal as a (non-deterministic) procedure call.  In a well-moded
program~\cite{Debray88jlc}, terms in input position are seen as the actual
arguments of this procedure, and terms in output position yield return
values.

In this section, we will not formalize the notion of well-modedness ---
see~\cite{Debray88jlc} for Prolog and~\cite{sarnat10thesis} for Twelf --- nor
refine our operational semantics to make goal evaluation order and unification
explicit --- see~\cite{pientka03thesis}.  We will instead refine our
compilation process to account for mode information and produce compiled
programs that, if well-moded, can be executed without appealing to
unification.

\subsection{Target Language}
\label{sec:pil2-target}


In $\LcB$, a (well-moded) Horn clause $\A{\vec{y}}{p\:\Ivec{t}\:\Ovec{t} \Pim
  a_1 \Pim \ldots \Pim a_n}$ was compiled into
$\A{\Ivec{x}\;\Ovec{x}}{(p\:\Ivec{x}\:\Ovec{x} \Pim
  \E{\vec{y}}{(\Eq{\Ivec{x}}{\Ivec{t}} \And \Eq{\Ovec{x}}{\Ovec{t}} \And a_1
    \And \ldots \And a_n)})}$.  Here, the left-to-right execution order forces
us to guess the final values of the output variables $\Ovec{x}$ before the
goals in its body have been fully executed.  In $\LcC$, we will move the
equality $\Eq{\Ovec{x}}{\Ovec{t}}$ after the last goal $a_n$.  Since
$\Ovec{x}$ appear nowhere else in the residual, this equality is no more than
an assignment of the computed instance of $\Ovec{t}$ to $\Ovec{x}$.
Accordingly, we will write it as $\Assg{\Ovec{x}}{\Ovec{t}}$.  Furthermore, in
a well-moded program, this clause will be invoked with ground terms in input
position, so that $\Ivec{x}$ will be bound to ground terms.  Then, the input
equality $\Eq{\Ivec{x}}{\Ivec{t}}$ will match the variables in $\Ivec{t}$ with
appropriate subterms.  For this reason, we will write it as
$\Match{\Ivec{x}}{\Ivec{t}}$.  Expanding each goal $a_i$ into
$q_i\;\Ivec{t}_i\;\Ovec{t}_i$, the above clause will be compiled (almost) as
follows, where the arrows represent the data flow of a well-moded execution
(note that it parallels the control flow):
\vspace{-5ex}%
$$
\newcommand{\PIM}{\hspace{0.8em}\Pim\hspace{0.8em}}%
\newcommand{\AND}{\hspace{0.8em}\And\hspace{0.8em}}%
\xymatrix@C=-0.4em@R=0.5em{
   &&&&&&&&&&&&&&&&&&&&&&
\\ &&&&&&&&&&&&&&&&&&&&&&
\\ \A{\Ivec{x}\:\Ovec{x}}{}(
   &  p&\Ivec{x} \ar@{<.}[uu]               \ar@(d,d)[rrrrr]
          &\Ovec{x} \ar@{.>}[uu]
& \PIM & (\E{\vec{y}}{}
&      & \Ivec{x} & \Match{}{} & \Ivec{t}   \ar@(d,d)[rrr]
& \AND & q_1
       & \Ivec{t}_1
       & \Ovec{t}_1                         \ar@(d,d)[rr]
& \AND & \ldot&\!\!\ldot\!\!&\ldot          \ar@(d,d)[rrr]
& \AND & q_n
       & \Ivec{t}_n
       & \Ovec{t}_n                         \ar@(d,d)[rrrr]
& \AND & \Ovec{x}                           \ar@(ul,ul)[llllllllllllllllllll]
                  & \Assg{}{}  & \Ovec{t}
& ))
\\\relax
}
$$
\medskip
\noindent

When executing an atomic goal, it is desirable to separate the call from the
verification that the output terms returned by the caller match the expected
output terms in this goal.  We will do so by rewriting any atomic goal
$q\:\Ivec{t}\:\Ovec{t}$ in a compiled clause into the formula
$\E{\vec{z}}{(q\:\Ivec{t}\:\vec{z} \And \Match{\vec{z}}{\Ovec{t}})}$ for fresh
variables $\vec{z}$.  This
transformation preserves the left-to-right control and data flow.
No special provision needs to be made for the input arguments of $q$ as
variables in it will have been instantiated to ground terms at the moment the
call is made.


\medskip

Next, we again generalize this intuition to any formula in $\Ls$, not just
Horn clauses.  Our third target language, $\LcC$, is defined by the following
grammar.

\medskip
\noindent
{\arraycolsep=0.25em
\newcommand{\sep}{\hspace{0.5em}|\hspace{0.5em}}
$\begin{array}[t]{@{}r@{\hspace{0.8em}} cr l@{}}
   \mbox{\emph{Goal Matches:}}
& M & ::= & \True
      \sep  M \And \Match{z}{\OT{t}}
\\ \mbox{\emph{Atomic Goals:}}
& F & ::= & p\;\Ivec{t}\;\Ovec{z} \And M
      \sep  \E{z}{F}
\\ \mbox{\emph{Goals:}}
& G & ::= & F
      \sep  C \Imp G
      \sep  \A{x}{G}
\\[1ex]
   \mbox{\emph{Clauses:}}
& C & ::= & R \Imp p\;\Ivec{x}\;\Ovec{x}
      \sep  \A{x}{C}
\\ \mbox{\emph{Residuals:}}
& R & ::= & \Match{\IN{x}}{t}  
      \sep  \Assg{\OT{x}}{t}
      \sep  \True
      \sep  R \And G
      \sep  \E{x}{R}
\end{array}
\hspace{-4em}
\begin{array}[t]{@{}r@{\hspace{0.8em}} cr l@{}}
  \mbox{\emph{Programs:}}
& \Psi & ::= & \cdot
       \sep  \Psi, C
\end{array}$}%

\medskip
\noindent
Residuals ($R$) refine the equality predicate $\Eq{x}{t}$ of $\LcB$ into a
matching predicate $\Match{x}{t}$ and an assignment predicate $\Assg{x}{t}$.
At our level of abstraction, they behave just like equality.  During
well-moded execution, the match predicate will have the form
$\Match{t_g}{t_v}$ where $t_g$ is a ground term while $t_v$ may contain
variables.  It will bind these variables to ground subterms of $t_g$, thereby
realizing matching.  However, presented with programs that are not well-moded,
the terms $t_g$ cannot be assumed to be ground and $\Match{}{}$ performs
unification.  The assignment predicate will be called as $\Assg{x}{t}$ where
$x$ is a variable and $t$ a term --- a ground term for well-moded programs.
It simply binds $x$ to $t$.  Compiled clauses and programs are just like in
$\LcB$.

Following the motivations above, an atomic goal $p\;\Ivec{t}\;\Ovec{t}$ is not
compiled any more to itself as in $\LcB$, but to a formula $F$ of the form
$\E{\vec{z}}{(q\:\Ivec{t}\:\Ovec{z} \And \Match{\Ivec{z}}{\Ovec{t}})}$.  In
the grammar above, we isolated the match predicates
$\Match{\Ivec{z}}{\Ovec{t}}$ as the non-terminal $M$.


\medskip

\begin{figure}[t]
\begin{center}
\leavevmode
\newcommand{\rsp}{1.8ex}
\newcommand{\rspl}{-.5ex}
\newcommand{\trule}[1][-1.5ex]{\rule[#1]{0em}{0ex}}
\newcommand{\hhfill}{\hfill}
\fboxsep=0pt

\noindent
\fbox{\scriptsize
$\begin{array}{@{\hfill}c@{\hfill}}
\makebox[\textwidth]{}\\[-2.5ex]
\multicolumn{1}{l}{\bf \scriptstyle Goals\;Matches\trule}
\\[-0.5ex]
  \cRule
   {}
   {\mResC{}{\True}}
   {m2\_true}
\hhfill
  \cRule
   {\mResC{}{M}}
   {\mResC{}{M \And \Match{t}{t}}}
   {m2\_mtch}
\\[\rspl]
\hline\\[-4.5ex]
\multicolumn{1}{l}{\bf \scriptstyle Atomic\;Goals\trule}
\\[-0.5ex]
  \cRule
   {\cResC[p\;\Ivec{t}\;\Ovec{t}]{\Psi, C, \Psi'}{C}
    \qq\qq
    \mResC{}{M}}
   {\fResC{\Psi, C, \Psi'}{p\;\Ivec{t}\;\Ovec{t} \And M}}
   {a2\_atm}
\hhfill
  \cRule
   {\fResC{\Psi}{[t/z]R}}
   {\fResC{\Psi}{\E{z}{R}}}
   {a2\_exists}
\\[\rspl]
\hline\\[-4.5ex]
\multicolumn{1}{l}{\bf \scriptstyle Goals\trule}
\\[-0.5ex]
  \cRule
   {\fResC{\Psi}{F}}
   {\gResC{\Psi}{F}}
   {g2\_f}
\hhfill
  \cRule
   {\gResC{\Psi, C}{G}}
   {\gResC{\Psi}{C \Imp G}}
   {g2\_imp}
\hhfill
  \cRule
   {c \; \mbox{\em ``new''}
    \qq
    \gResC{\Psi}{[c/x]G}}
   {\gResC{\Psi}{\A{x}{G}}}
   {g2\_all}
\\[\rspl]
\hline\hline\\[-4.5ex]
\multicolumn{1}{l}{\bf \scriptstyle Clauses\trule}
\\[-1.0ex]
  \cRule
   {\rResC{\Psi}{R}}
   {\cResC{\Psi}{R \Imp a}}
   {c2\_imp}
\hhfill
  \cRule
   {\cResC{\Psi}{[t/x]C}}
   {\cResC{\Psi}{\A{x}{C}}}
   {c2\_all}
\\[\rspl]
\hline\\[-4.5ex]
\multicolumn{1}{l}{\bf \scriptstyle Residuals\trule}
\\[-1.0ex]
  \cRule
   {}
   {\gResC{\Psi}{\Match{t}{t}}}
   {r2\_mtch}
\hhfill
  \cRule
   {}
   {\gResC{\Psi}{\Assg{t}{t}}}
   {r2\_assg}
\hhfill
  \cRule
   {}
   {\rResC{\Psi}{\True}}
   {r2\_true}
\\[\rsp]
  \cRule
   {\rResC{\Psi}{R}
    \qq
    \gResC{\Psi}{G}}
   {\rResC{\Psi}{R \And G}}
   {r2\_and}
\hhfill
  \cRule
   {\rResC{\Psi}{[t/x]R}}
   {\rResC{\Psi}{\E{x}{R}}}
   {r2\_exists}
\\[-1ex]\relax
\end{array}$}

\caption{Search Semantics of $\LcC$.}
\label{fig:pil2-resolution}
\end{center}
\end{figure}

We specify the operational semantics of $\LcC$ by means of the following five
judgments:
$$
\begin{array}{l@{\hspace{1.5em}}p{18em}}
   \mResC{\phantom{\Psi}}{M}
 & \emph{$M$ is provable}
\\ \fResC{\Psi}{F}
 & \emph{$F$ is uniformly provable from $\Psi$}
\\ \gResC{\Psi}{G}
 & \emph{$G$ is uniformly provable from $\Psi$}
\\ \cResC{\Psi}{C}
 & \emph{$a$ is immediately entailed by $C$ in $\Psi$}
\\ \rResC{\Psi}{R}
 & \emph{$R$ is uniformly provable from $\Psi$}
\end{array}
$$
which parallel the grammar just presented.  The resulting operational
semantics is shown in Figure~\ref{fig:pil2-resolution}.  The rules for clauses
are unchanged with respect to $\LcB$ while that language's residual rule for
equality has been duplicated into isomorphic rules for matching and
assignment.  The rules for compiled goals have instead proliferated due to our
handling of terms in output position in atomic goals.  Observe that rule
\rname{a2\_atm} is essentially a combination of rule \rname{g1\_atm} in $\LcB$
and the rule for conjunction.  Rules \rname{a2\_exists} and \rname{m2\_true}
are just the standard rules for existential quantification and truth.  Rule
\rname{m2\_mtch} combines the rules for conjunction and matching.

\medskip

Just like in the case of $\LcB$, the rules in Figure~\ref{fig:pil2-resolution}
construct proofs that are uniform~\cite{Miller91apal}, which makes $\LcC$ an
abstract logic programming language.  In a successful derivation, this
operational semantics decomposes a goal to formulas of the form $F =
\E{\vec{z}}{(p\:\Ivec{t}\:\Ovec{z} \And \Match{\Ivec{z}}{\Ovec{t}})}$ (rules
in the ``Goals'' segment).  Then, rules \rname{a2\_exists}, \rname{m2\_mtch}
and \rname{m2\_true} necessarily reduce it in a few steps into the atomic
formula $p\:\Ivec{t}\:\Ovec{t}$.  Similarly to $\LcB$, the left premise of
rule \rname{a2\_atm} selects a clause and focuses on it until it finds a
potentially matching head (``Clauses'' segment).  It then proceeds to
decomposing its body (``Residuals'' segment) and the cycle repeats with
whatever goals it finds in there.

As just noticed, any atomic goal $F$ of the form
$\E{\Ovec{z}}{(p\:\Ivec{t}\:\Ovec{z} \And \Match{\Ovec{z}}{\Ovec{t}})}$ is
necessarily reduced to $p\:\Ivec{t}\:\Ovec{t}$ by as many applications of rule
\rname{a2\_exists} as there are variables in $\Ovec{z}$, a pass-through
instance of \rname{a2\_atm} via its right branch, and a similar number of uses
of rules \rname{m2\_mtch} and \rname{m2\_true} respectively.  This entails
that the macro-rule \rname{a2\_atm'}, on the left-hand side of the following
display, is derivable:
$$
\cRule
 {\gResC{\Psi}{p\:\IN{t}\:\OT{t}}}
 {\gResC{\Psi}
   {\E{\vec{z}}{(p\:\Ivec{t}\:\vec{z} \And
       \Match{\vec{z}}{\Ovec{t}})}}}
 {a2\_atm'}
\hspace{6em}
\cRule
 {\cResC[p\;\Ivec{t}\;\Ovec{t}]{\Psi, C, \Psi'}{C}}
 {\fResC{\Psi, C, \Psi'}{p\;\Ivec{t}\;\Ovec{t}}}
 {a2\_atm''}
$$
Having factored
rule \rname{a2\_atm'} out, the work performed by \rname{a2\_atm} degenerates
to rule \rname{a2\_atm''} on the right-hand side of the above display,
which is akin to \rname{u\_atm}.  The system obtained by replacing the
\rname{m2\_*} and \rname{a2\_*} rules as well as \rname{g2\_f} with rules
\rname{a2\_atm'} and \rname{a2\_atm''} is indeed equivalent to the rule set in
Figure~\ref{fig:pil2-resolution}.

Rule \rname{a2\_atm'} entices us to interpret the compiled formula
$\E{\vec{z}}{(p\:\Ivec{t}\:\vec{z} \And \Match{\vec{z}}{\Ovec{t}})}$ for an
atomic goal $p\:\IN{t}\:\OT{t}$ as a synthetic operator
$\ms{call}\:p\:\Ivec{t} \Match{}{} \Ovec{t}$ which invokes a clause for $p$
with its (ground) input arguments $\Ivec{t}$ and matches the returned values
against its terms $\Ovec{t}$ in output position.

Having recovered atomic goals $p\:\Ivec{t}\:\Ovec{t}$ through rules
\rname{a2\_atm'} and \rname{a2\_atm''}, we can carry out a sequence of
reasoning steps similar to what led us to the backchaining rule for $\LcB$.
Exposing the trailing assignments, a generic compiled clause $C$ has the form
$\A{\Ivec{x}\:\Ovec{x}}{(\E{\vec{y}}{R \And \Ovec{x} \Assg{}{} \Ovec{s})} \Imp
  p\:\Ivec{x}\:\Ovec{x}}$.  In a successful derivation, all rule
\rname{a2\_atm''} does is to pick such a clause.  Then, applications of rule
\rname{c2\_all} will instantiate variables $\Ivec{x}\:\Ovec{x}$ with the terms
$\Ivec{t}\:\Ovec{t}$, and next rule \rname{c2\_imp} will invoke the instantiated
residual $[\Ivec{t}/\Ivec{x},\Ovec{t}/\Ovec{x}](\E{\vec{y}}{R \And \Ovec{x}
  \Assg{}{} \Ovec{s}})$.  Now, because $\Ovec{x}$ does not occur in $R$ and
$\Ivec{x}\:\Ovec{x}$ cannot appear in $\Ovec{s}$, this formula reduces to
$\E{\vec{y}}{([\Ivec{t}/\Ivec{x}]R \And \Ovec{t} \Assg{}{} \Ovec{s})}$ by
pushing the substitution in.  Rule \rname{r2\_exists} will then instantiate the
variables $\vec{y}$ with terms $\vec{u}$ (which cannot mention variables
$\Ivec{x}\:\Ovec{x}$).  Pushing this substitution in yields the formula
$[\Ivec{t}/\Ivec{x},\vec{u}/\vec{y}]R \And \Ovec{t} \Assg{}{}
[\vec{u}/\vec{y}]\Ovec{s}$ since variables in $\vec{y}$ can occur in neither
$\Ivec{t}$ nor $\Ovec{t}$.  Finally, by rule \rname{r2\_assg}, $\Ovec{t}$ and
$[\vec{u}/\vec{y}]\Ovec{s}$ must be equal in a successful derivation.  This
necessary sequence of steps is captured by the following derived backchaining
macro-rule,
$$
\cRule
 {\rResC
   {\Psi, C, \Psi'}
 {[\Ivec{t}/\Ivec{x},\vec{u}/\vec{y}]R}}
 {\gResC
   {\Psi,
     \underbrace{\A{\Ivec{x}\:\Ovec{x}}{
                   (\E{\vec{y}}{R \And \Ovec{x} \Assg{}{} \Ovec{s})}
                 \Imp p\:\Ivec{x}\:\Ovec{x}}}_C,
    \Psi'}
   {p\;\Ivec{t}\;[\vec{u}/\vec{y}]\Ovec{s}}}
 {g2\_atm'}
$$
where we have carried out the assignment $\Ovec{t} \Assg{}{}
[\vec{u}/\vec{y}]\Ovec{s}$ in the conclusion.  This rule can be seen as a
refinement of \rname{g1\_atm'} in $\LcB$ that makes use of the trailing
assignment in the compiled clauses of $\LcC$.  With this derived inference,
rules \rname{a2\_atm''}, \rname{c2\_imp} and \rname{c2\_all} become
unnecessary: the system consisting of rules \rname{a2\_atm'},
\rname{g2\_atm'}, the goal rules for implication and universal quantification,
and the residual rules is equivalent to that in
Figure~\ref{fig:pil2-resolution}.

Taking rule \rname{g2\_atm'} as primitive amounts to replacing compiled clauses
with the following synthetic connective, which refines $\LcB$'s $\Lam_p
\vec{x}.\, R$.
$$
\begin{array}{cccl}
\underbrace{\A{\Ivec{x}\:\Ovec{x}}{}
p\:\Ivec{x}\:\Ovec{x}
\subset}
& \E{\vec{y}}{(
     R
& \And & \underbrace{\Ovec{x} \Assg{}{} \Ovec{t}}
)}
\\
\Lam_p \Ivec{x}.\,
& \E{\vec{y}}{}(
     R
&    ; & \ms{return}\:\Ovec{t})
\end{array}
$$
The variables $\vec{y}$ are then interpreted as local variables for the
execution of this clause.  In this, they are akin to the \verb"Y"$n$ permanent
variables of the WAM~\cite{Aitkaci91book}.

In a valid proof in this system, an occurrence of \rname{a2\_atm'} is always
immediately followed by an instance of \rname{g2\_atm'}: the conclusion of the
latter must match the premise of the former.  This fact realizes the
requirement that, upon returning from a call, the output terms, here
$[\vec{u}/\vec{y}]\Ovec{s}$, must be checked against the terms in output
position of the caller.





\subsection{Compilation}
\label{sec:pil2-compilation}

Compilation transforms logic programs in $\Ls$ to compiled programs in
$\LcC$.  The input does not have to be well-moded at the level of detail
considered here, but this would be operationally advantageous in a refinement
of the semantics in Figure~\ref{fig:pil2-resolution} that handles quantifiers
lazily.  We will make use of two auxiliary notions in this section:
pseudo clauses that we encountered already in
Section~\ref{sec:pil1-compilation} and the analogous notion of pseudo atomic
goal.  They are defined as follows:

\medskip
\noindent
{\arraycolsep=0.25em
\newcommand{\sep}{\hspace{0.5em}|\hspace{0.5em}}
$\begin{array}{@{}r@{\hspace{0.8em}} cr l@{}}
   \mbox{\emph{Pseudo Clauses:}}
&  \CC & ::= & \Box \Imp p\:\vec{x}
       \sep  \A{x}{\CC}
\\ \mbox{\emph{Pseudo Atomic Goals:}}
&  \GG & ::= & p\:\Ivec{t}\:\Ovec{z} \And \Box
       \sep  \E{z}{\GG}
\end{array}$}%

\medskip
\noindent
Just like pseudo clauses retain the outer structure of a clause replacing the
embedded residual with a hole ($\Box$), pseudo atomic goals have a hole in
place of their trailing matches.  The general form of pseudo clauses and
pseudo atomic formulas, accounting for input and output positions, are
$\A{\Ivec{x}\:\Ovec{x}}{\Box \Imp p\:\Ivec{x'}\:\Ovec{x'}}$ and
$\E{\Ovec{z}}{(p\:\Ivec{t}\:\Ovec{z'} \And \Box)}$.
%
In Section~\ref{sec:pil1-compilation}, wrote $\CC[R]$ for the replacement of
the hole of $\CC$ with the residual $R$ and noted that variable capture could
(and generally will) occur.  Similarly, we write $\GG[M]$ for replacement of
the hole of $\GG$ with matches $M$.

\begin{figure}[t]
\begin{center}
\leavevmode
\newcommand{\rsp}{1.8ex}
\newcommand{\rspl}{-.5ex}
\newcommand{\trule}[1][-1.5ex]{\rule[#1]{0em}{0ex}}
\newcommand{\hhfill}{\hfill}
\fboxsep=0pt

\noindent
\fbox{\scriptsize
$\begin{array}{@{\hfill}c@{\hfill}}
\makebox[\textwidth]{}\\[-2.5ex]
\multicolumn{1}{l}{\bf \scriptstyle Programs\trule}
\\[-0.5ex]
  \cRule
   {}
   {\cmpProgC{\cdot}{\cdot}}
   {p2c\_empty}
\hhfill
  \cRule
   {\cmpProgC{\G}{\Psi}
    \qq
    \cmpClauseC[\CC]{A}{R}{O}}
   {\cmpProgC{\G, A}{\Psi, \CC[R \And O]}}
   {p2c\_clause}
\\[\rspl]
\hline\\[-4.5ex]
\multicolumn{1}{l}{\bf \scriptstyle Heads\trule}
\\[-1.0ex]
  \cRule
   {}
   {\cmpHeadC{\vec{x}}{p}{\Box \Imp p\:\vec{x}}{\True}{\True}}
   {h2c\_p}
\\[\rsp]
  \cRule
   {\cmpHeadC{x\;\vec{x}}{a}{\CC}{I}{O}
    \qq
    x \: \mbox{\em ``new''}}
   {\cmpHeadC{\vec{x}}{a\;\IN{t}}{\A{x}{\CC}}{I \And \Match{x}{\IN{t}}}{O}}
   {h2c\_in}
\hhfill
  \cRule
   {\cmpHeadC{x\;\vec{x}}{a}{\CC}{I}{O}
    \qq
    x \: \mbox{\em ``new''}}
   {\cmpHeadC{\vec{x}}{a\;\OT{t}}{\A{x}{\CC}}{I}{\Assg{x}{\OT{t}} \And O}}
   {h2c\_ot}
\\[\rspl]
\hline\\[-4.5ex]
\multicolumn{1}{l}{\bf \scriptstyle Clauses\trule}
\\[0.0ex]
  \cRule
   {\cmpHeadC{\cdot}{a}{\CC}{I}{O}}
   {\cmpClauseC[\CC]{a}{I}{O}}
   {c2c\_atm}
\hhfill
  \cRule
   {\cmpGoalC{A}{G}
    \qq
    \cmpClauseC[\CC]{B}{R}{O}}
   {\cmpClauseC[\CC]{A \Imp B}{R \And G}{O}}
   {c2c\_imp}
\hhfill
  \cRule
   {\cmpClauseC[\CC]{A}{R}{O}}
   {\cmpClauseC[\CC]{\A{x}{A}}{\E{x}{R}}{O}}
   {c2c\_all}
\\[\rspl]
\hline\\[-4.5ex]
\multicolumn{1}{l}{\bf \scriptstyle Atomic\:goals\trule}
\\[-1.0ex]
  \cRule
   {}
   {\cmpAtmC{\vec{t}}{p}{p\:\vec{t} \And \Box}{\True}}
   {a2c\_p}
\hhfill
  \cRule
   {\cmpAtmC{\IN{t}\:\vec{t}}{a}{\GG}{M}}
   {\cmpAtmC{\vec{t}}{a\;\IN{t}}{\GG}{M}}
   {a2c\_in}
\hhfill
  \cRule
   {\cmpAtmC{\vec{t}\:z}{a}{\CC}{M}
    \qq
    z \: \mbox{\em ``new''}}
   {\cmpAtmC{\vec{t}}{a\;\OT{t}}{\E{z}{\GG}}{\Match{z}{\OT{t}} \And M}}
   {a2c\_ot}
\\[\rspl]
\hline\\[-4.5ex]
\multicolumn{1}{l}{\bf \scriptstyle Goals\trule}
\\[-0.0ex]
  \cRule
   {\cmpAtmC{\cdot}{a}{\GG}{M}}
   {\cmpGoalC{a}{\GG[M]}}
   {g2c\_atm}
\hhfill
  \cRule
   {\cmpClauseC[\CC]{A}{R}{O}
    \qq
    \cmpGoalC{B}{G}}
   {\cmpGoalC{A \Imp B}{\CC[R \And O] \Imp G}}
   {g2c\_imp}
\hhfill
  \cRule
   {\cmpGoalC{A}{C}}
   {\cmpGoalC{\A{x}{A}}{\A{x}{C}}}
   {g2c\_all}
\\[-1ex]\relax
\end{array}$}

\caption{Compilation of $\Ls$ into $\LcC$.}
\label{fig:pil2-compilation}
\end{center}
\end{figure}

The compilation process is modeled by the following five judgments, which are
reminiscent of the compilation judgments $\LcB$.  They are more complex
because clause compilation now needs to handle both matching and assignment as
opposed to a generic equality.  Furthermore, a new judgment is needed to
compile atomic goals.
$$
\begin{array}{l@{\hspace{1.5em}}p{20em}}
   \cmpProgC{\G}{\Psi}
 & \emph{Program $\G$ is compiled to $\Psi$}
\\ \cmpHeadC{\vec{x}}{a}{\CC}{I}{O}
 & \emph{Head $a$ with $\vec{x}$ is compiled to $\CC$, $I$ and $O$}
\\ \cmpClauseC{A}{R}{O}
 & \emph{Clause $A$ is compiled to $\CC$, $R$ and $O$}
\\ \cmpAtmC{\vec{t}}{a}{\GG}{M}
 & \emph{Atomic goal $a$ with $\vec{t}$ is compiled to $\GG$ and $M$}
\\ \cmpGoalC{A}{G}
 & \emph{Goal $A$ is compiled to $G$}
\end{array}
$$
We write $I$ and $O$ for a conjunction of matches (compilation of terms in
input position) and assignments (compilation of output terms), respectively,
in the body of a compiled clause.  In compiled atomic goals, we write $M$ for
a conjunction of matches.

The rules for compilation, which define these judgments, are shown in
Figure~\ref{fig:pil2-compilation}.  Compiling a clause $A$, modeled by the
judgment $\cmpClauseC{A}{R}{O}$, returns a pseudo clause $\CC$, the residual
$R$ (inclusive of input matches) and the output assignments $O$ that will fill
its hole.  The rules in the ``Clauses'' segment build up this residual
starting with the compilation of its head, which is displayed in the ``Heads''
segment.  The rules therein differ from the similar inference for $\LcB$ by
the fact that they dispatch terms in input and output positions in the $I$ and
$O$ zones of the judgment as matches and assignments respectively.  Residuals
and assignments are plugged in the hole of the pseudo clause once this clause
has been fully compiled, as can be seen in the ``Programs'' segment and in
rule \rname{g2c\_imp}.

The compilation of goals differs from $\LcB$ for the treatment of atomic
formulas: upon encountering an atom $a$, the compilation appeals to the new judgment
$\cmpAtmC{\cdot}{a}{\GG}{M}$.  It generates a pseudo atomic formula $\GG$ and
matches $M$, which are integrated in rule \rname{g2c\_atm}.  The zone to the
left of the turnstile serves as an accumulator, very much like when compiling heads.

Target language, $\LcC$, is sound and complete with respect to $\Ls$.  The
following lemma collects some auxiliary results needed to prove this property.
The first two statements are proved by induction on the structure of $a$; the
third by induction on the given derivation.

\noindent{\parbox{\linewidth}{%
\begin{lemma}
\begin{itemize}
\itemsep=0pt
\item%
  If $\cmpHeadC{\vec{x}}{a}{\CC}{I}{O}$, then for any term sequence $\vec{t}$
  of the same length as $\vec{x}$ and program $\Psi$ we have
  $\cResC[a\:\vec{t}]{\Psi}{[\vec{t}/\vec{x}](\CC[I \And O])}$.
\item%
  If $\cmpAtmC{\vec{t}}{a}{\GG}{M}$, then for all $\Psi$ we have
  $\cResC[a\:\vec{t}]{\Psi}{\GG[M]}$.
\item%
  If $\cResC{\Psi}{\CC[R]}$, then $\rResB{\Psi}{R}$.
\end{itemize}
\end{lemma}}

We have the following soundness and completeness theorems for $\LcC$.  In both
cases, the proof proceeds by mutual induction over the first derivation in the
antecedent.

\noindent{\parbox{\linewidth}{%
\begin{theorem}[Soundness of the compilation to $\LcC$]
\label{th:pil2-sound}
\begin{itemize}
\itemsep=0ex
\item %
    If   \ $\unid{\G}{A}$,
         \ $\cmpProgC{\G}{\Psi}$
  \ and  \ $\cmpGoalC{A}{G}$,
  \ then \ $\gResC{\Psi}{G}$.
\item %
    If   \ $\immd{\G}{A}$,
         \ $\cmpProgC{\G}{\Psi}$
  \ and  \ $\cmpClauseC[\CC]{A}{R}{O}$,
  \ then \ $\cResC{\Psi}{\CC[R \And O]}$.
\end{itemize}
\end{theorem}}

\noindent{\parbox{\linewidth}{%
\begin{theorem}[Completeness of the compilation to $\LcC$]
\label{th:pil2-complete}
\begin{itemize}
\itemsep=0ex
\item %
    If   \ $\gResC{\Psi}{G}$,
         \ $\cmpProgC{\G}{\Psi}$
  \ and  \ $\cmpGoalC{A}{G}$,
  \ then \ $\unid{\G}{A}$.
\item %
    If   \ $\cResC{\Psi}{C}$,
         \ $\cmpProgC{\G}{\Psi}$,
         \ $C = \CC[R \And O]$
  \ and  \ $\cmpClauseC[\CC]{A}{R}{O}$,
  \ then \ $\immd{\G}{A}$.
\end{itemize}
\end{theorem}}

\begin{figure}[t]
\fbox{%
\hspace*{-1em}%
\parbox{\linewidth}{%
\begin{enumerate}\small
\item%
$\begin{array}[t]{@{}l@{\hspace{3.5em}}c@{\hspace{2em}}l@{}}
  \begin{array}[t]{@{}ll@{}}
     \multicolumn{2}{@{}l@{}}{
       \A{E_1}{}\A{E_2}{}
       \A{T_1}{}\A{T_2}{}}
  \\      & \ms{of}\;(\ms{app}\;E_1\;E_2)\;T_2
  \\
  \\
  \\ \Pim & \ms{of}\;E_1\;(\ms{arr}\;T_1\;T_2)
  \\ \Pim & \ms{of}\;E_2\;T_1
  \end{array}
& \raisebox{-10ex}{ \ $\cmpProgB{}{}$ \hspace*{-0.8em} }
& \begin{array}[t]{@{}ll@{}l@{\;}l@{}}
     \multicolumn{4}{@{}l@{}}{\A{x_1}{}\A{x_2}{}}
  \\      & \multicolumn{3}{@{}l@{}}{\ms{of}\;x_1\;x_2}
  \\ \Pim &(&\multicolumn{2}{@{}l@{}}{
          \E{E_1}{}\E{E_2}{}
          \E{T_1}{}\E{T_2}{\True}}
  \\ && \And & \Match{x_1}{\ms{app}\;E_1\;E_2}
  \\ && \And & \E{z_1}{(\ms{of}\;E_1\;z_1 \And \Match{z_1}{\ms{arr}\;T_1\;T_2}
                                          \And \True)}
  \\ && \And & \E{z_2}{(\ms{of}\;E_2\;z_2 \And \Match{z_2}{T_1} \And \True)}
  \\ && \And & \Assg{x_2}{T_2} \And \True)
  \end{array}
\end{array}$


\medskip%
\item%
$\begin{array}[t]{@{}lcl@{}}
  \begin{array}[t]{@{}ll@{}}
     \multicolumn{2}{@{}l@{}}{
       \A{E}{}
       \A{T_1}{}\A{T_2}{}}
  \\      & \ms{of}\;(\ms{lam}\;T_1\;E)\;(\ms{arr}\;T_1\;T_2)
  \\
  \\
  \\ \Pim & (\A{x}{} 
  \\
  \\      & \hspace{2em} \ms{of}\;x\;T_1
  \\
  \\      & \hspace{0.5em} \Imp \ms{of}\;(E\;x)\;T_2)
  \end{array}
& \raisebox{-14.5ex}{$\cmpProgB{}{}$}
& \begin{array}[t]{@{}ll@{}l@{\;}l@{}}
     \multicolumn{4}{@{}l@{}}{\A{x_1}{}\A{x_2}{}}
  \\      & \multicolumn{3}{@{}l@{}}{\ms{of}\;x_1\;x_2}
  \\ \Pim &(&\multicolumn{2}{@{}l@{}}{
          \E{E}{}
          \E{T_1}{}\E{T_2}{\True}}
  \\ &&\And & \Match{x_1}{\ms{lam}\;T_1\;E}
  \\ &&\And & \E{z}{}((\A{x}{} (\begin{array}[t]{@{}l@{\;}l@{}}
                             & \begin{array}[t]{@{}l@{\;}l@{}}
                                  \multicolumn{2}{@{}l@{}}{\A{x_1'}{}\A{x_2'}{\True}}
                               \\ \And & \Match{x_1'}{x}
                               \\ \And & \ms{of}\;x_1'\;x_2'
                               \\ \And & \Assg{x_2'}{T_1} \And \True)
                               \end{array}
                     \\ \Imp & \ms{of}\;(E\;x)\;z)
                     \end{array}
  \\ &&     & \ \ \ \ \ \And \Match{z}{T_2 \And \True})
  \\ &&\And & \Assg{x_2}{\ms{arr}\;T_1\;T_2} \And \True)
  \end{array}
\end{array}$
\end{enumerate}}}
\caption{$\LcC$ Compilation Example}
\label{fig:pl2-example}
\end{figure}

To conclude this section, we revisit our ongoing examples.  Here, we assume
that the mode of the predicate $\ms{of}$ is $\ms{of}\;\IN{\:}\;\OT{\:}$ ---
the first argument is input and the second output.  The result of compiling
our two familiar clauses into $\LcC$ is shown in Figure~\ref{fig:pl2-example}.
As in Section~\ref{sec:pil1-compilation}, the moded compilation process offers
ample opportunities for optimization: matches and assignments with variables
on both side and the corresponding existential quantification can often be
elided, and all occurrences of $\True$ can be optimized away.

It is instructive to rewrite these clauses with the two synthetic connectives
introduced earlier for $\LcC$, again omitting $\True$ for readability:
$$
\begin{array}{lcl}
    \Lam_{\ms{of}}\; x_1.
  & \multicolumn{2}{l}{\E{E_1}{}\E{E_2}{}\E{T_1}{}\E{T_2}{}
     \;\;   \Match{x_1}{\ms{app}\;E_1\;E_2}}
\\&  \And & \ms{call}\: (\ms{of}\;E_1) \;\Match{}{}\; (\ms{arr}\;T_1\;T_2)
   \;\And\; \ms{call}\: (\ms{of}\;E_2) \;\Match{}{}\; T_1 ;
\\&       & \ms{return}\: T_2
\\[1ex]
    \Lam_{\ms{of}}\; x_1.
  & \multicolumn{2}{l}{\E{E}{}\E{T_1}{}\E{T_2}{}
     \;\;   \Match{x_1}{\ms{lam}\;T_1\;E}}
\\&  \And & \A{x}{}(\Lam_{\ms{of}}\; x_1'.\;\;
                                 \Match{x_1'}{x}\;;
                                 \ms{return}\: T_1)
                     \;\Imp\; \ms{call}\: (\ms{of}\;(E\;x)) \;\Match{}{}\; T_2;
\\&       & \ms{return}\: (\ms{arr}\;T_1\;T_2)
\end{array}
$$

\section{Larger Source Languages}
\label{sec:larger-languages}

In~\cite{Cervesato98jicslp}, we illustrated our original abstract logical
compilation method on the language of hereditary Harrop formulas.  This
language differs from $\Ls$ for the presence of conjunction (formulas of the
form $A \And B$) and truth ($\True$).  While our original treatment could
handle them easily (in a clause position, they were compiled to disjunctions
and falsehood respectively), the approach taken in Sections~\ref{sec:pil1}
and~\ref{sec:pil2} does not support them directly.  The problem is that, as soon
as we allow these connectives, clauses can have multiple heads (or even none).
Consider for example:
$$
\A{x}{}\A{y}{}
q\:x\:y \Imp (p_1\:x\:y \And (r\:x\;y \Imp p_2\:x))
$$
This clause has two heads: $p_1\:x\:y$ and $p_2\:x$.  What should it be
compiled to?  To ensure immediacy (embodied in the macro-rule
\rname{g1\_atm'}), our compilation strategy produces a pseudo clause applied to
a residual, thereby exposing the (flattened) head of a compiled clause as close
to the top level as possible.  How to achieve this now that there may be more
than one head?

One approach to dealing with this problem is to observe that $\sAnd$
distributes over (the antecedent of) $\sImp$ and $\sA$.  By doing so to the
above example, we obtain the formula
$$
(\A{x}{}\A{y}{}
q\:x\:y \Imp p_1\:x\:y)
\And
(\A{x}{}\A{y}{}
q\:x\:y \Imp r\:x\;y \Imp p_2\:x)
$$
Observe that it is a conjunction of $\Ls$ clauses.  Each of them can now be
compiled as in Section~\ref{sec:pil1} and the results can be combined by means
of a disjunction.  This approach generalizes to the full language of
hereditary Harrop formulas.  It pushes the conjunctions to the outside,
leaving inner formulas resembling the clauses of $\Lc$ (conjunction and truth
in a goal position are left alone as they are not problematic).  Clauses with
no head (e.g., $A \Imp \True$) are reduced to $\True$.  These preprocessing
steps can be implemented as a source-code transformation or integrated in the
compilation process.

The other abstract logic programming language examined
in~\cite{Cervesato98jicslp} is the language of linear hereditary Harrop
formulas, found at the core of Lolli~\cite{Hodas94ic} and LLF~\cite{ic02}.
The improved compilation process discussed in this paper extends directly in
the presence of linearity.  Because linear hereditary Harrop formulas feature
a form of conjunction and truth, the technical device just outlined is needed
to obtain workable compiled clauses.

\section{Future Work}
\label{sec:future}

The discussion in Section~\ref{sec:pil2} sets the stage for a nearly functional
operational semantics of well-moded programs.  Indeed, given an atomic goal
with ground terms in its input positions, proof search will instantiate its
output positions to ground terms, if it succeeds.  Being in a logic
programming setting, more than one answer could be returned.  Indeed, for
well-moded programs, the clauses for a predicate implement a partial,
non-deterministic function.  This observation informed the choice of the
notation for the synthetic operators we exposed: $\ms{call}\:p\:\Ivec{t}
\Match{}{} \Ovec{t}$ and $\Lam_p \Ivec{x}.\,\E{\vec{y}}{}(R;
\ms{return}\:\Ovec{t})$.

Now we believe that, in the case of well-moded programs, a more detailed
operational semantics that exposes variable manipulations using logical
variables and explicit substitutions (and restricts the execution order) can
bring this functional interpretation to the surface.  This would provide a
logical justification for the natural impulse to give well-moded programs a
semantics that is typical of functional programming languages, where atomic
predicates carry just input terms and from which the terms in output position
emerge by a process of reduction.

In future work, we intend to carry out this program by giving such a detailed
operational semantics to $\Ls$ as well as well-moding rules.  The goal will
then be to perform logical transformations, akin to what we did in this paper,
that expose this functional semantics for well-moded programs.  It would also
allow us to prove formally that the operator $\Match{}{}$ of
Section~\ref{sec:pil2} can indeed be implemented as matching rather than
general unification.

\section*{Acknowledgments}
This work was supported by the Qatar National Research Fund under grant NPRP
09-1107-1-168.  We are grateful to Frank Pfenning, Carsten Sch\"urmann, Robert
J. Simmons and Jorge Sacchini for the many fruitful discussions, as well as to
the anonymous reviewers.

\label{lastpage}
\bibliography{better-compilation}

\begin{thebibliography}{}

\bibitem[\protect\citeauthoryear{A{\"\i}t-Kaci}{A{\"\i}t-Kaci}{1991}]{Aitkaci9%
1book}
{\sc A{\"\i}t-Kaci, H.} 1991.
\newblock {\em Warren's Abstract Machine: a Tutorial Reconstruction}.
\newblock MIT Press.

\bibitem[\protect\citeauthoryear{B{\"o}rger and Rosenzweig}{B{\"o}rger and
  Rosenzweig}{1995}]{Boerger95fmpa}
{\sc B{\"o}rger, E.} {\sc and} {\sc Rosenzweig, D.} 1995.
\newblock The {WAM} --- definition and compiler correctness.
\newblock In {\em Logic Programming: Formal Methods and Practical
  Applications}, {C.~Beierle} {and} {L.~Pluemer}, Eds. Computer Science and
  Artificial Intelligence, vol.~11. North-Holland, 21--90.

\bibitem[\protect\citeauthoryear{Cervesato}{Cervesato}{1998}]{Cervesato98jicsl%
p}
{\sc Cervesato, I.} 1998.
\newblock {Proof-Theoretic Foundation of Compilation in Logic Programming
  Languages}.
\newblock In {\em 1998 Joint International Conference and Symposium on Logic
  Programming --- JICSLP'98}, {J.~Jaffar}, Ed. MIT Press, Manchester, UK,
  115--129.

\bibitem[\protect\citeauthoryear{Cervesato and Pfenning}{Cervesato and
  Pfenning}{2002}]{ic02}
{\sc Cervesato, I.} {\sc and} {\sc Pfenning, F.} 2002.
\newblock {A Linear Logical Framework}.
\newblock {\em Information \& Computation\/}~{\em 179,\/}~1, 19--75.

\bibitem[\protect\citeauthoryear{Cervesato, Pfenning, Walker, and
  Watkins}{Cervesato et~al\mbox{.}}{2003}]{cmu-cs-02-102}
{\sc Cervesato, I.}, {\sc Pfenning, F.}, {\sc Walker, D.}, {\sc and} {\sc
  Watkins, K.} 2003.
\newblock {A Concurrent Logical Framework II: Examples and Applications}.
\newblock Technical Report CMU-CS-02-102, Department of Computer Science,
  Carnegie Mellon University, Pittsburgh, PA. March 2002, revised May.

\bibitem[\protect\citeauthoryear{Debray and Warren}{Debray and
  Warren}{1988}]{Debray88jlc}
{\sc Debray, S.~K.} {\sc and} {\sc Warren, D.~S.} 1988.
\newblock Automatic mode inference for logic programs.
\newblock {\em Journal of Logic Programming\/}~{\em 5}, 207--229.

\bibitem[\protect\citeauthoryear{Hodas and Miller}{Hodas and
  Miller}{1994}]{Hodas94ic}
{\sc Hodas, J.~S.} {\sc and} {\sc Miller, D.} 1994.
\newblock Logic programming in a fragment of intuitionistic linear logic.
\newblock {\em Information and Computation\/}~{\em 110,\/}~2, 327--365.

\bibitem[\protect\citeauthoryear{Jaffar, Michaylov, Stuckey, and Yap}{Jaffar
  et~al\mbox{.}}{1992}]{Jaffar92pldi}
{\sc Jaffar, J.}, {\sc Michaylov, S.}, {\sc Stuckey, P.}, {\sc and} {\sc Yap,
  R.} 1992.
\newblock An abstract machine for \emph{CLP$({\cal R})$}.
\newblock In {\em Proceedings of the SIGPLAN'92 Conference on Programming
  Language Design and Implementation --- PLDI'92}. San Francisco, CA.

\bibitem[\protect\citeauthoryear{Miller and Nadathur}{Miller and
  Nadathur}{1986}]{Miller86iclp}
{\sc Miller, D.} {\sc and} {\sc Nadathur, G.} 1986.
\newblock Higher-order logic programming.
\newblock In {\em Proceedings of the Third International Logic Programming
  Conference}, {E.~Shapiro}, Ed. London, 448--462.

\bibitem[\protect\citeauthoryear{Miller, Nadathur, Pfenning, and
  Scedrov}{Miller et~al\mbox{.}}{1991}]{Miller91apal}
{\sc Miller, D.}, {\sc Nadathur, G.}, {\sc Pfenning, F.}, {\sc and} {\sc
  Scedrov, A.} 1991.
\newblock Uniform proofs as a foundation for logic programming.
\newblock {\em Annals of Pure and Applied Logic\/}~{\em 51}, 125--157.

\bibitem[\protect\citeauthoryear{Nadathur and Mitchell}{Nadathur and
  Mitchell}{1999}]{nadathur99cade}
{\sc Nadathur, G.} {\sc and} {\sc Mitchell, D.~J.} 1999.
\newblock System description: Teyjus --- a compiler and abstract machine based
  implementation of lambda prolog.
\newblock In {\em Sixteenth Conference on Automated Deduction (CADE'99)},
  {H.~Ganzinger}, Ed. 287--291.

\bibitem[\protect\citeauthoryear{Pfenning and Sch{\"u}rmann}{Pfenning and
  Sch{\"u}rmann}{1999}]{Pfenning99cade}
{\sc Pfenning, F.} {\sc and} {\sc Sch{\"u}rmann, C.} 1999.
\newblock {System Description: Twelf --- A Meta-Logical Framework for Deductive
  Systems}.
\newblock In {\em Proceedings of the 16th International Conference on Automated
  Deduction --- CADE-16}. Springer-Verlag LNAI 1632, Trento, Italy, 202--206.

\bibitem[\protect\citeauthoryear{Pientka}{Pientka}{2003}]{pientka03thesis}
{\sc Pientka, B.} 2003.
\newblock Tabled higher-order logic programming.
\newblock Ph.D. thesis, Department of Computer Science, Carnegie Mellon
  University.

\bibitem[\protect\citeauthoryear{Russinoff}{Russinoff}{1992}]{Russinoff92jlp}
{\sc Russinoff, D.~M.} 1992.
\newblock A verified {P}rolog compiler for the {W}arren abstract machine.
\newblock {\em Journal of Logic Programming\/}~{\em 13}, 367--412.

\bibitem[\protect\citeauthoryear{Sarnat}{Sarnat}{2010}]{sarnat10thesis}
{\sc Sarnat, J.} 2010.
\newblock Syntactic finitism in the metatheory of programming languages.
\newblock Ph.D. thesis, Department of Computer Science, Yale University.

\bibitem[\protect\citeauthoryear{Stirling}{Stirling}{2009}]{stirling09lmcs}
{\sc Stirling, C.} 2009.
\newblock Decidability of higher-order matching.
\newblock {\em Logical Methods in Computer Science\/}~{\em 5,\/}~3.

\bibitem[\protect\citeauthoryear{Warren}{Warren}{1983}]{Warren83tr}
{\sc Warren, D. H.~D.} 1983.
\newblock An abstract {P}rolog instruction set.
\newblock Technical Note 309, SRI International, Menlo Park, CA. Oct.

\bibitem[\protect\citeauthoryear{Watkins, Cervesato, Pfenning, and
  Walker}{Watkins et~al\mbox{.}}{2003}]{cmu-cs-02-101}
{\sc Watkins, K.}, {\sc Cervesato, I.}, {\sc Pfenning, F.}, {\sc and} {\sc
  Walker, D.} 2003.
\newblock {A Concurrent Logical Framework I: Judgments and Properties}.
\newblock Technical Report CMU-CS-02-101, Department of Computer Science,
  Carnegie Mellon University, Pittsburgh, PA. March 2002, revised May.

\end{thebibliography}

\end{document}